\documentclass[12pt,a4paper]{article}

\usepackage{psfig}

\newcommand{\bd}{\begin{displaymath}}
\newcommand{\ed}{\end{displaymath}}
\newcommand{\be}{\begin{equation}}
\newcommand{\ee}{\end{equation}}
\newcommand{\beq}{\begin{eqnarray}}
\newcommand{\eeq}{\end{eqnarray}}  
\newcommand{\beqs}{\begin{eqnarray*}}
\newcommand{\eeqs}{\end{eqnarray*}}
  
\begin{document}

\title{Perturbative expansion for the half-integer rectilinear disclination line in the Landau-de Gennes theory }

\author{H. Arod\'{z}\thanks{Marian Smoluchowski Institute of Physics, Jagiellonian University, Reymonta 4, 30-059 Cracow, 
Poland.}        \hspace*{0.5cm}
and   \hspace*{0.5cm}
R. Pe\l ka\thanks{H. Niewodnicza\'{n}ski Institute of Nuclear Physics, Radzikowskiego 152, 31-342 Cracow, Poland.}} 

\maketitle

\begin{abstract}
\noindent The structure of the half-integer rectilinear disclination line within the framework of the Landau-de Gennes 
effective theory of nematic liquid crystals is investigated. The consistent perturbative expansion is constructed for the 
case of $L_2\neq 0$. It turns out that such expansion can be performed around only a discrete subset of an infinite set 
of the degenerate zeroth order solutions. These solutions correspond to the positive and negative wedge disclination lines 
and to four configurations of the twist disclination line. The first order corrections to both the order parameter field 
as well as the free energy of the disclination lines have been found. The results for the free energy are compared
with the ones obtained in the Frank-Oseen-Zocher director description.  
\end{abstract}

\vspace*{4cm}

\noindent
PACS: 61.30.Jf, 11.27.+d \\
Preprint TPJU-17/2002

\newpage

\section{Introduction}
\label{intro}

Disclination lines in nematic liquid crystals rank among the most popular topological defects encountered in condensed 
matter physics \cite{chan,degennes,chaikin,chandra}. This notwithstanding, there are still a number of interesting 
theoretical problems to be solved, which particularly holds for the Landau-de Gennes theory, for its review and discussion 
see, \textit{e.g.}, \cite{gramsbergen}. Such a state of affairs is partly due to the relative complexity of the order 
parameter structure in that theory. Not less importantly, the Landau-de Gennes model appeared a decade later than the 
Frank-Oseen-Z\"{o}cher director formalism \cite{degennes1}. In contrast to that formalism, which leaves the question of the core structure of the disclination line open, the relevant solution in the Landau-de Gennes model is smooth everywhere \cite{lyuk,schop,meiboom,disc1}. Although both the formalisms account for the intrinsic elastic anisotropy of a nematic medium, only in the director formalism the exact solution for generic values of the Frank elastic constants had been found \cite{dzyalosh,anisimov}. At the same time, to the best of our knowledge, even approximate analytical results for the case of $L_2\neq 0$ in the Landau-de Gennes model, which correspond to the two-constant approximation in the director formalism,  are still missing. The only results are those reported by Schopohl and Sluckin in \cite{schop}, where numerical calculations have been carried out for half-integer wedge disclinations. 

In this paper we consider a generic half-integer disclination line within the framework of the Landau-de Gennes theory. Our main result is the formulation of a consistent perturbative expansion for the case of $L_2\neq 0$. It must be mentioned that we draw extensively on the results obtained in \cite{disc1}. Therein the isotropic $L_2=0$ case has been thoroughly discussed, which represents the zeroth order of the perturbative scheme constructed here. Via some symmetry considerations we find that only around a restricted set of the zeroth order solutions a perturbative expansion can be consistently performed. The presented scheme allows one to generate the corrections in all consecutive orders of the expansion. The forms of the first order contributions to the order parameter have been found explicitly. For the positive and negative wedge disclination lines in the simplest high symmetry case of $\beta=\frac{1}{3}$ accounted for in \cite{disc1} also numerical calculations have been performed. What is important, the relevant corrections proved to be at least one order of magnitude smaller than the zeroth order contribution, which validates the applicability of the preturbative approach. Next, the polynomial approximation suggested in \cite{disc1} has been taken full advantage of in order to estimate the first order corrections to the free energy. In accord with the results obtained within the two-constant approximation in the director formalism \cite{anisimov}, for $L_2>0$ ($L_2<0$) the twist disclination line (the positive wedge disclination line) has been found to have the lowest free energy. The first order correction to the free energy of the twist disclination line was found to be logarithmically divergent, whereas that of both the wedge disclination lines turned out to be finite. The corresponding first order corrections to the value of the 'critical' temperature below which the smooth disclination line is energetically favourable have been estimated. The corresponding 'critical' temperatures have been found to increase (decrease) for $L_2>0$ ($L_2<0$) for all types of the disclination line.   

The paper is organized as follows: in Sec.\ref{prel} the preliminary considerations have been presented, Sec.\ref{cons} includes the very construction of the perturbative expansion, whereas Sec.\ref{fecor} focuses on the first order corrections to the free energy and to the value of the 'critical' temperature, in Sec.\ref{fr} some final remarks have been made, and  in Appendix the derivations of the first order corrections to the order parameter are to be found.    

\section{Preliminaries}\label{prel}

\subsection{The Landau-de Gennes theory}\label{ldgt}

The order parameter for the nematic liquid crystal in the Landau-de Gennes theory is a symmetric traceless real tensor ${Q}_{ij}$, where $i, j=1,2,3$ and the corresponding free energy density is given by the following formulae

\be
\mathcal{F}=\frac{1}{2}L_1\partial_{i}Q_{jk}\partial_{i}Q_{jk}+\frac{1}{2}L_2\partial_{i}Q_{ik}\partial_{j}Q_{jk}+V(\hat{Q}), \label{1}
\ee

\noindent where

\be
V(\hat{Q})=-\frac{a}{2}\mathrm{Tr}\left(\hat{Q}^2\right)-\frac{b}{3}\mathrm{Tr}\left(\hat{Q}^3\right)+\frac{c}{4}\left(\mathrm{Tr}\left(\hat{Q}^2\right)\right)^2. \label{2}
\ee

For the standard stability reasons constants $a$ and $c$ are assumed to be positive. Furthermore, for most of the explored nematogenic compounds, see \textit{e.g.} \cite{vertog}, the ground state proved to be that of a uniaxial ordering, that is why we set also the constant $b$ to be positive. Simple stability analysis for basic inhomogeneous conformations of the order parameter \cite{vertog} shows that the elastic constant $L_1$ should be positive, whereas for $L_2$ the following restriction has to be satisfied 

\be
L_2>-\frac{3}{2}L_1. \label{3}
\ee

\noindent Hence, negative values of that constant are not excluded.

The order parameter in the ground state is a constant tensor $\hat{Q}_g$, whose form we obtain from the necessary condition for local extrema of the free energy. It is given by the following formula

\be
\hat{Q}_g=\mathcal{O}\hat{Q}_0\mathcal{O}^{\mathrm{T}}, \label{4} 
\ee

\noindent where 

\be
\hat{Q}_0=\eta_0\left(\begin{array}{ccc}
2 & 0 & 0 \\
0 & -1 & 0 \\
0 & 0 & -1
\end{array}\right), \label{5}
\ee

\noindent and

\be
\eta_0=\frac{b+\sqrt{b^2+24ac}}{12c},\label{6}
\ee

\noindent whereas the transformations $\mathcal{O}$ without loss of generality can be restricted to the set of the proper three-dimensional rotations $SO(3)$. Thus the vacuum manifold is continuously degenerate. 

In \cite{disc1} it has been found that for the rectilinear disclination line in the Landau-de Gennes theory with $L_2=0$ the order parameter takes on the following axially symmetric form

\be
\hat{Q}_d=\frac{3}{2}\eta_0\mathcal{O}(\phi)\left(\begin{array}{ccc}
\frac{S(\rho)}{3}+R(\rho) & 0 & 0 \\
0 & \frac{S(\rho)}{3}-R(\rho) & 0 \\
0 & 0 & -\frac{2}{3}S(\rho)
\end{array}\right)\mathcal{O}^{\mathrm{T}}(\phi), \label{7}
\ee

\noindent where $\eta_0$ is given by the formula (\ref{6}), functions $S(\rho), R(\rho)$ are a pair of smooth structure functions satisfying the following boundary conditions

\be
S(0)=\mathrm{const}, \hspace{1.5 cm} S(\infty)=1, \label{8}
\ee

\be
R(0)=0, \hspace{1.5 cm} R(\infty)=1, \label{9}
\ee

\noindent and the matrix $\mathcal{O}(\phi)$ has the form

\be
\mathcal{O}(\phi)=\left(\begin{array}{ccc}
\cos\frac{\phi}{2} & - \sin\frac{\phi}{2} & 0 \\
\sin\frac{\phi}{2} & \cos\frac{\phi}{2} & 0 \\
0 & 0 & 1 
\end{array}\right). \label{10}
\ee

\noindent The core of the disclination line given by the above solution is smooth with a planar uniaxial phase on the line, the maximal eigenvalue of $\hat{Q}$ being negative, reached via the ring of biaxiality as one approaches the line from the far-off area filled with the uniaxial phase. On taking some simple assumptions as to the creation of the singular disclination core \cite{disc1}, it was pointed out that for sufficiently low temperatures parametrized by   

\be
\beta=\frac{b\eta_0}{3a}\label{10a}
\ee

\noindent the smooth disclination line represents an energetically preferred conformation, wheras for $\beta$ large enough a singular core with the isotropic phase inside will be more favourable. The 'critical' value of $\beta$ dividing those two stability domains was estimated to be equal to $0.12$.  
 
\subsection{Degeneracy of the ground state of a rectilinear disclination line}\label{degen}

It is clear that \textit{Ansatz} (\ref{7}) does not exhaust the set of the possible solutions of the pertinent Euler-Lagrange equation for the rectilinear $m=\frac{1}{2}$ dislination line when $L_2=0$. Other topologically equivalent solutions are given by the formula

\be
\hat{Q}(\rho,\phi)=\mathcal{O}_1\hat{Q}_{d}(\rho,\phi)\mathcal{O}^{T}_{1},\label{11}
\ee

\noindent where $\mathcal{O}_1\in SO(3)$.

\noindent The topological equivalence does not yet mean the physical one. While all the solutions (\ref{11}) are characterized by the same free energy, they describe distinguishable spatial configurations of the nematogenic molecules; \textit{e.g.} $\hat{Q}_d$ corresponds to a configuration where far off the disclination line all the molecules lie on avarage in the $xy$-plane (a wedge disclination line), whereas inserting for $\mathcal{O}_1$ in formula (\ref{11}) the matrix

\be
\left(\begin{array}{ccc}
1 & 0 & 0 \\
0 & 0 & -1 \\
0 & 1 & 0 
\end{array}\right),\label{12}
\ee

\noindent one arrives at the configuration referred to in the literature \cite{degennes} as a twist dislination line where all the far-off molecules tend to lie in the plane parallel to the direction of the line. On the other hand, among solutions (\ref{11}) one may find such which, failing to display mathematical identity, result in the same spatial configurations of molecules, the difference originating from the mere change of the frame of reference. In general, two solutions $\hat{Q}$ and $\hat{\tilde{Q}}$ give physically equivalent spatial configurations, if there exists such a global coordinate transformation $\mathcal{O}_c\in H_d$ that

\be
\mathcal{O}_c\hat{Q}\mathcal{O}^{\mathrm{T}}_c=\hat{\tilde{Q}}. \label{13}
\ee

\noindent Group $H_d$ contains all proper orthogonal transformations which leave the position of the disclination line unchanged. We purposefully omit the group of one-dimensional translations along the line as irrelevant here. Thus $H_d=SO(2)$. Eq. (\ref{13}) means that physically nonequivalent configurations correspond to the different right cosets $SO(3)/SO(2)$. Knowing that any rotation can be parametrized with the set of three Euler angles, one finds easily that the looked for elements of the coset are given by the formula

\be
\mathcal{O}_1(\Theta,\Psi)=\mathcal{O}_y(\Theta)\mathcal{O}_z(\Psi), \label{14}
\ee

\noindent where $\mathcal{O}_y, \mathcal{O}_z$ refer to the rotations about the $y$- and $z$-axis, respectively, and $\Theta\in[0,\pi],\ \Psi\in[0,2\pi)$. One should also exclude the case where both the rotation matrices in the above formula commute, \textit{i.e.} when $\theta=0$ or $\pi$, and different values of angle $\Psi$ lead to the physically indistinguishable configurations. Apart from that, there exists one more symmetry in the system. Namely, one can easily check that 

\be
\hat{Q}_{d}=\mathcal{O}_{0}\hat{Q}_{d}\mathcal{O}_{0}^{T},\label{15}
\ee

\noindent where 

\be
\mathcal{O}_{0}=\left(\begin{array}{ccc}
-1 & 0 & 0 \\
0 & -1 & 0 \\
0 & 0 & 1 \end{array}\right).\label{16}
\ee   

\noindent  This means that transformations $\mathcal{O}_1(\Theta,\Psi)$ i $\mathcal{O}_1(\Theta,\Psi)\mathcal{O}_{0}$ result in mathematically equivalent solutions. Moreover, because  

\be
\mathcal{O}_1(\Theta,\Psi)\mathcal{O}_{0}=\mathcal{O}_1(\Theta,\pi+\Psi),\label{17}
\ee
 
\noindent we get nonequivalent solutions for $\Psi$ in the reduced segment of $[0,\pi)$. Summing it all up, the physically nonequivalent configurations are given by formula (\ref{11}), where $\mathcal{O}_1$ is given by formula (\ref{14}) with pairs $(\Theta,\Psi)$ lying in the interior of a square $(0,\pi)\times (0,\pi)$ or on its boundary segment $[0,\pi]\times \{0\}$.

\section{Construction of the perturbative expansion}\label{cons}

\subsection{First steps}\label{fsteps}

The main goal of this section consists in reporting the construction of the perturbative scheme in the case of $L_2\neq 0$. The unperturbed system corresponds to the so called one-constant approximation in the director formalism, where $K_1=K_2=K_3=9\eta_0^2(2L_1+ L_2)$, whereas the perturbation takes on the following form
 
\be
\delta\mathcal{F}=\frac{1}{2}L_2\left((\mathrm{div}\hat{Q})^2-\frac{1}{2}(\mathrm{grad}\hat{Q})^2\right), \label{18}
\ee

\noindent where $(\mathrm{div}\hat{Q})^2=\partial_l Q_{lj}\partial_k Q_{kj}$, and $(\mathrm{grad}\hat{Q})^2=\partial_k Q_{jl}\partial_k Q_{jl}$.  As the expansion parameter we identify the dimensionless ratio $\varepsilon=2L_2/(2L_1+L_2)$. One can easily check that the following identity holds:

\be
\varepsilon=\frac{2(\bar{K}-K_2)}{\bar{K}},\label{19}
\ee 

\noindent where $\bar{K}=(K_1+K_3)/2$. Using the experimental data \cite{dejeu} for two popular representatives of nematic liquid crystals (for MBBA at transition temperature $\approx 320\ K$, $K_1=7.1\times 10^{-12}\ N$, $K_2=4.0\times 10^{-12}\ N$, $K_3=9.2\times 10^{-12}\ N$; for DIBAB at transition temperature $\approx 305\ K$, $K_1=4.8\times 10^{-12}\ N$, $K_2=3.0\times 10^{-12}\ N$, $K_3=4.7\times 10^{-12}\ N$), we arrive at the following estimations

\bd
\varepsilon_{\mathrm{MBBA}}\approx1.01,\hspace{2 cm}\varepsilon_{\mathrm{DIBAB}}\approx0.74.
\ed

\noindent The fact that these values of $\varepsilon$ are large does not mean that the perturbative calculations are out of place here, as one should remember that what is relevant is not the value of the expansion parameter itself but the order of magnitude of the full consecutive corrections. It will be demonstrated later that the first order corrections are at least one order of magnitude smaller than the zeroth order contributions, so the perturbative approach finds quite a good justification.

For the sake of operational convenience let us introduce a rescaled order parameter $\hat{q}$ defined by the following formula

\be
\hat{Q}=\frac{3}{2}\eta_0\hat{q},\label{20}
\ee

\noindent as well as dimensionless coordinates defined as follows

\be
\begin{array}{cc}
x^{i}=\tilde{\xi}_0 s^{i}, \ \ \ i=1,2,3, & \mathrm{gdzie} \ \ \ \tilde{\xi}_0=\sqrt{\frac{2L_1+L_2}{3a}}\end{array}.\label{21}
\ee 

\noindent In the variables introduced above the Euler-Lagrange equation takes on the following form 

\be
\begin{array}{c}
\tilde{\partial}_k\tilde{\partial}_k q_{ij}-\frac{1}{6}\varepsilon\left[2(\tilde{\partial}_k\tilde{\partial}_l q_{kl})\delta_{ij}-3\left(\tilde{\partial}_i\tilde{\partial}_k q_{kj}+\tilde{\partial}_j\tilde{\partial}_k q_{ki}\right)+3\tilde{\partial}_k\tilde{\partial}_k q_{ij}\right]+\frac{2}{3}q_{ij}+3\beta(\hat{q}^2)_{ij}- \\
-\frac{1}{4}(1+3\beta)Tr(\hat{q}^2)q_{ij}-\beta Tr(\hat{q}^2)\delta_{ij}=0,
\end{array}\label{22}
\ee 

\noindent where the derivatives with respect to the dimensionless coordinates are marked out by a tilde. We seek its solutions assuming the perturbative expansion

\be
\hat{q}=\hat{q}_0+\varepsilon\hat{q}_1+\varepsilon^2\hat{q}_2 + \ldots.\label{23}
\ee
 
Inserting expansion (\ref{23}) into equation (\ref{22}), and equating to zero the factors appearing at the subsequent powers of $\varepsilon$ we find equations governing the dynamics of the subsequent corrections to the order parameter. 

In the zeroth order we obtain the equation discussed in detail in \cite{disc1} 
 
\be
\tilde{\partial}_k\tilde{\partial}_k \hat{q}_0+\frac{2}{3}\hat{q}_0+3\beta\hat{q}_0^2-\frac{1}{4}(1+3\beta)Tr(\hat{q}_0^2)\hat{q}_0-\beta Tr(\hat{q}_0^2)I=0.\label{24}
\ee 

\noindent According to what has been said in the previous section the physically nonequivalent solutions of equation (\ref{24}) are parametrized by a pair of angles  $(\Theta, \Psi)$ with values in $(0,\pi)\times(0,\pi)\cup[0,\pi]\times\{0\}$. They have the form

\be
\hat{q}_{0}^{(\Theta,\Psi)}(s,\phi)=\mathcal{O}_1(\Theta,\Psi)\hat{q}_d(s,\phi)\mathcal{O}_1^{T}(\Theta,\Psi),\label{25}
\ee

\noindent where $\mathcal{O}_1(\Theta,\Psi)$ is given by the formula (\ref{14}) and $\hat{q}_d(s,\phi)$ has the following form 

\be
\hat{q}_d(s,\phi)=\left(\begin{array}{ccc}
\frac{1}{3}S(s)+R(s)\cos\phi & R(s)\sin\phi & 0 \\
R(s)\sin\phi & \frac{1}{3}S(s) -R(s)\cos\phi & 0 \\
0 & 0 & -\frac{2}{3} S(s) 
\end{array}\right).\label{26}
\ee
      
\noindent The equations in higher orders of the perturbative expansion are of the general form

\be
\hat{\mathcal{L}}\hat{q}_k=\hat{\mathcal{N}}_k,\label{27}
\ee

\noindent where $k=1, 2, \ldots $, $\hat{\mathcal{L}}$ denotes a linear operator in the space $\mathcal{M}$ of coordinate-dependent symmetric traceless and real matrices $3\times 3$. Its action on a generic element $\hat{m}=\hat{m}(s^i)$ in that space is defined by the formula

\be
\hat{\mathcal{L}}\hat{m}=\tilde{\partial}_k\tilde{\partial}_k\hat{m}+\frac{2}{3}\hat{m}+3\beta(\hat{q}_0\hat{m}+\hat{m}\hat{q}_0)-\frac{1}{2}(1+3\beta) Tr(\hat{q}_0\hat{m})\hat{q}_0-\frac{1}{4}(1+3\beta) Tr(\hat{q}_0^2)\hat{m}-2\beta Tr(\hat{q}_0\hat{m})I.\label{28}    
\ee

\noindent The consecutive terms on the r.h.s. of Eq. (\ref{27}) can be obtained from the formal recipe

\be
\hat{\mathcal{N}}_k=\frac{1}{k!}\left.\frac{\partial^k\hat{h}}{\partial\varepsilon^k}\right|_{\varepsilon=0},\label{29}
\ee

\noindent where matrix $\hat{h}$ is defined by the formula 

\be
\begin{array}{c}
h_{ij}(s^{i};\varepsilon)=\frac{1}{6}\varepsilon[2(\tilde{\partial}_k\tilde{\partial}_k q_{kl})\delta_{ij}-3(\tilde{\partial}_i\tilde{\partial}_k q_{kj}+\tilde{\partial}_j\tilde{\partial}_k q_{ki})+3\tilde{\partial}_k\tilde{\partial}_k q_{ij}]-3\beta(\hat{q}^2-\hat{q}_0\hat{q}-\hat{q}\hat{q}_0)_{ij}+ \\
+\frac{1}{4}(1+3\beta)\left[Tr(\hat{q}^2)\hat{q}-2Tr(\hat{q}_0\hat{q})\hat{q}_0-Tr(\hat{q}_0^2)\hat{q}\right]_{ij}+\beta Tr[\hat{q}(\hat{q}-2\hat{q}_0)]\delta_{ij},
\end{array}\label{30}
\ee

\noindent with $\hat{q}$ in the form of expansion (\ref{23}). For example, 

\beq
(\hat{\mathcal{N}}_1)_{ij}&=&\frac{1}{6}[2(\tilde{\partial}_k\tilde{\partial}_l q_{0kl})\delta_{ij}-3(\tilde{\partial}_i\tilde{\partial}_k q_{0kj}+\tilde{\partial}_j\tilde{\partial}_k q_{0ki})+3\tilde{\partial}_k\tilde{\partial}_k q_{0ij}]\label{31} \\
(\hat{\mathcal{N}}_2)_{ij}&=&\frac{1}{6}[2(\tilde{\partial}_k\tilde{\partial}_l q_{1kl})\delta_{ij}-3(\tilde{\partial}_i\tilde{\partial}_k q_{1kj}+\tilde{\partial}_j\tilde{\partial}_k q_{1ki})+3\tilde{\partial}_k\tilde{\partial}_k q_{1ij}]-3\beta(\hat{q}_1^2)_{ij} \nonumber \\
&+& \frac{1}{4}(1+3\beta)\left[Tr(\hat{q}_1^2)\hat{q}_0+2Tr(\hat{q}_0\hat{q}_1)\hat{q}_1\right]_{ij}+\beta Tr(\hat{q}_1^2)\delta_{ij}.\label{32}
\eeq

$\hat{\mathcal{L}}$ and $\hat{\mathcal{N}}_k$ depend on the zeroth order solution (\ref{29},\ref{30}). In spite of that, we do not fit those entities with an additional subscript referring to a given $\hat{q}_0^{(\Theta,\Psi)}$ but stipulate that all the formulae to be found hereafter refer to a generic choice of a pair $(\Theta, \Psi)$, unless stated otherwise. 

\subsection{Symmetry considerations}\label{symcon}

It is clear that the unperturbed system is invariant under the action of the full $SO(3)$ group, \textit{i.e.} the free energy is the same for any solution of the form

\be
\hat{q}_0^{\prime}=\mathcal{O}\hat{q}_0\mathcal{O}^{T},\label{33}
\ee

\noindent where $\mathcal{O}\in SO(3)$. A simple consequence of this symmetry is the existence of the corresponding rotational zero modes. One can easily check that they have the form

\be
\hat{\theta}_{\alpha}=[\hat{t}_{\alpha},\hat{q}_0],\label{34}
\ee

\noindent where $\hat{t}_{\alpha}$ denotes the generator of the rotation about axis $\alpha$ ($\alpha=1,2,3$). The modes (\ref{34}) are eigenfunctions of operator $\hat{\mathcal{L}}$ belonging to the eigenvalue equal to zero. 

Yet, the $SO(3)$ symmetry of internal rotations is not the only one in the system. The other one is the group of translations along the three independent directions. This symmetry originates from the homogeneity of the system, as we assume here that all the material constants $a,b,c,L_1,L_2$ do not depend on the coordinates $x^i$. Similarly, there exist three corresponding translational zero modes 
 
\be
\hat{\tau}_i=\frac{\partial\hat{q}_0}{\partial s^i},\label{35}
\ee

\noindent where $i=1,2,3$. Furthermore, because $\hat{q}_0$ does not depend on variable $s^3$, the mode $\hat{\tau}_3$ corresponding to the trivial translations along the disclination line vanishes. On differentiating both sides of equation (\ref{24}) with respect to $s^i$ ($i=1,2$) we see that these modes are the eigenfunctions of $\hat{\mathcal{L}}$ belonging to the eigenvalue $0$. 

The zero modes discussed above possess a clear physical interpretation, \textit{i.e.} they appear as the consequence of the symmetry of the unperturbed model.

Note that 

\be
[\hat{\mathcal{L}},\tilde{\partial}_3]=0 \ \ \ \ \ \mathrm{i}\ \ \ \ \  \tilde{\partial}_3\hat{\mathcal{N}}_1=0. \label{37}
\ee
 
\noindent From that follows that there exists such a solution of equation (\ref{27}) with $k=1$ that is independent of coordinate $s^3$. However, the general solution of that equation does not exclude the $s^3$-dependence. That is, a particular solution dispalying the translational symmetry along the direction of the line may in general be appended by a solution breaking that symmetry. The latter one would have to be an eigenfunction of $\hat{\mathcal{L}}$ to the eigenvalue $0$ and would correspond to the excitations of the disclination line disturbing its rectilinearity. As the discussion of such perturbations will clearly exceed the intended scope of the present paper, we set that part of the full solution to zero in all consecutive orders of the perturbative expansion. Now the solution depends exclusivly on coordinates $s^1$, $s^2$. So, everywhere hereupon the Roman indices take on only two values $\{1,2\}$.

Now that we have explored the symmetries of the unperturbed model we may pass on to discussing their consequences. Let $\hat{\mu}$ denote any of the zero modes mentioned above. Next, multiplying both sides of Eq. (\ref{27}) by $\hat{\mu}$, taking the trace, and integrating the resulting equation over the circular area $\mathbf{C}(O,L)$ centered at the origin of the coordinate system, whose finite radius we denote by $L$, we arrive at the following equation

\be
\int_{\mathbf{C}(O, L)}d^2\vec{s}\ \mathrm{Tr}(\hat{\mu}\hat{\mathcal{L}}\hat{q}_k)=\int_{\mathbf{C}(O, L)}d^2\vec{s}\ \mathrm{Tr}(\hat{\mu}\hat{\mathcal{N}}_k).\label{38}
\ee 

\noindent The l.h.s. of the above equation can be rewritten into the form

\be
\mathrm{l.h.s.}=\int_{\mathbf{C}(O, L)}d^2\vec{s}\ \mathrm{Tr}[\hat{\mu}(\tilde{\partial}_l\tilde{\partial}_l\hat{q}_k)-(\tilde{\partial}_l\tilde{\partial}_l\hat{\mu})\hat{q}_k+\hat{q}_k\hat{\mathcal{L}}\hat{\mu}].\label{39}
\ee

\noindent The contribution from the last term in brackets vanishes, as $\hat{\mu}$ represents a zero mode, whereas the contribution from the remaining terms, on applying the Stokes theorem and some simple algebra, may be transformed into the following form 

\be
\mathrm{l.h.s.}=\int_{0}^{2\pi}d\phi\ \mathrm{Tr}\left.\left[\hat{\mu}\left(s\frac{\partial\hat{q}_k}{\partial s}\right)-\left(s\frac{\partial\hat{\mu}}{\partial s}\right)\hat{q}_k\right]\right|_{s=L},\label{40}
\ee

\noindent where $s=\sqrt{(s^1)^2+(s^2)^2}$. Here we arrive at the crucial point of this paper. Firstly, note that the rotational zero modes $\theta_{\alpha}$ reach their $\phi$-dependent asymptotic form as quick as $s^{-2}$, whereas the translational ones vanish as $s^{-1}$, see formulae (\ref{26}), (\ref{34}-\ref{35}). Secondly, it is only natural to assume that the consecutive corrections $\hat{q}_k$ are finite at infinity. That given, we conclude that the expression (\ref{40}) vanishes in the limit $L\rightarrow\infty$. As a consequence we obtain the following consistency conditions

\be
\lim_{L\rightarrow\infty}\int_{\mathbf{C}(O, L)}d^2\vec{s}\ \mathrm{Tr}(\hat{\mu}\hat{\mathcal{N}}_k)=0,\label{41}
\ee

\noindent where $k=1,2,\ldots $. They are to play an essential role in the construction.

Let us analyse the consistency conditions (\ref{41}) for $k=1$. To this end it is convenient to write $\hat{\mathcal{N}}_1$ in the form

\be
\hat{\mathcal{N}}_1=\frac{1}{6}\left[2\mathrm{Tr}(\hat{D}_0)I-3(\hat{D}_0+\hat{D}_0^{T})+3\tilde{\Delta}\hat{q}_0\right],\label{42}
\ee 

\noindent where $\hat{D}_0$ denotes the following matrix

\be
(\hat{D}_0)_{ij}=\tilde{\partial}_i\tilde{\partial}_l\left(\hat{q}_{0}\right)_{lj},\label{43}
\ee 

\noindent and $\tilde{\Delta}$ is the Laplace operator in the plane $(s^1,s^2)$. Remembering that the zero modes are symmetric and traceless matrices, the consistency condition can be rewritten as follows

\be
\lim_{L\rightarrow\infty}\int_{\mathbf{C}(O, L)}d^2\vec{s}\ \mathrm{Tr}\left[\hat{\mu}\left(2\hat{D}_0-\tilde{\Delta}\hat{q}_0\right)\right]=0.\label{44}
\ee 

To begin with, let us consider the translational zero modes $\hat{\tau}_{i}$ ($i=1,2$). Integrating by parts one can transform the l.h.s. of the condition (\ref{44}) into the following form 

\beq
& &\lim_{L\rightarrow\infty}\left\{\int_{\mathbf{\partial C}(O, L)} \left[2\delta_{kn}(\hat{\tau}_{i})_{jl}(\hat{\tau}_{n})_{kj}-(\hat{\tau}_{i})_{jk}(\hat{\tau}_{l})_{kj}\right]d\tilde{\sigma}^l \right.\nonumber \\
 & -&\left.\int_{\mathbf{C}(O, L)}d^2\vec{s}\ \tilde{\partial}_i\left[\tilde{\partial}_l(\hat{q}_0)_{lj}\tilde{\partial}_{k}(\hat{q}_0)_{kj}-\frac{1}{2}\tilde{\partial}_k(\hat{q}_0)_{jl}\tilde{\partial}_k(\hat{q}_0)_{lj}\right]\right\}.\label{45}
\eeq

\noindent As the expression within the brackets in the second integral vanishes in the limit $s\rightarrow\infty$, this integral will give no contribution. Neither will the surface integral, which follows immediately from the asymptotic behaviour of $\hat{\tau}_{i}$'s. Thus, the consistency conditions with the translational modes are trivial; they do not impose any restrictions on the zeroth order solution. There is nothing strange about it because the unperturbed model as well as the perturbation itself do not break the translational symmetry, so this symmetry holds for the full system. As a consequence, we obtain an infinite set of degenerate solutions which differ from each other in the position of the disclination core.

The consistency conditions with the rotational zero modes $\hat{\theta}_{\alpha}$ ($\alpha=1,2,3$), see Eqs. (\ref{34}, \ref{41}), can be rewritten as follows

\be
\lim_{L\rightarrow\infty}\int_{\mathbf{C}(O, L)}d^2\vec{s}\ \mathrm{Tr}\left\{\hat{t}_{\alpha}\left(2[\hat{q}_0,\hat{D}_0]-[\hat{q}_0,\tilde{\Delta}\hat{q}_0]\right)\right\}=0,\label{46}
\ee  

\noindent where the brackets denote the commutator. One can easily see that $[\hat{q}_0,\tilde{\Delta}\hat{q}_0]=0$. For the remaining term, on inserting the explicit form of $\hat{q}$ (Eq. (\ref{25})) and using polar coordinates in the plane $(s^1,s^2)$, we obtain

\beq
& & \lim_{L\rightarrow\infty}\left\{-\frac{\pi}{2}\sin^3\Theta\sin(4\Psi)\int_{0}^{L}ds s R\left(R^{\prime\prime}+\frac{R^{\prime}}{s}-\frac{R}{s^2}\right)\right\}=0,\label{47} \\
& & \lim_{L\rightarrow\infty}\left\{-\frac{2\pi}{3}\sin\Theta\cos\Theta\int_{0}^{L}ds s\left[S\left(S^{\prime\prime}+\frac{1}{s}S^{\prime}\right)-3R\left(R^{\prime\prime}+\frac{R^{\prime}}{s}-\frac{R}{s^2}\right)\right]\right\}=0,\label{48} \\
& & 0=0,\label{49}
\eeq  

\noindent for $\alpha=1,2,3$, respectively, where $'$ denotes the differentiation with respect to $s$. The integrals in (\ref{47}) and (\ref{48}) are nonzero and finite for every finite radius $L$, whereas in the limit $L\rightarrow\infty$ they diverge. Therefore, the only way to satisfy the consistency conditions is to select these pairs $(\Theta,\Psi)$ that solve the system of equations

\be
\left\{\begin{array}{c}
\sin^3\Theta\sin(4\Psi)=0, \\
\sin\Theta\cos\Theta=0,
\end{array}\right.\label{50}
\ee
  
\noindent which has the following solutions

\beq
& &\Theta=0, \ \ \ \Psi=0,\label{51} \\
& &\Theta=\pi, \ \ \ \Psi=0,\label{52} \\
& &\Theta=\frac{\pi}{2}, \ \ \ \Psi=0,\label{53} \\
& &\Theta=\frac{\pi}{2}, \ \ \ \Psi=\frac{\pi}{4},\label{54} \\
& &\Theta=\frac{\pi}{2}, \ \ \ \Psi=\frac{\pi}{2},\label{55} \\
& &\Theta=\frac{\pi}{2}, \ \ \ \Psi=\frac{3\pi}{4}.\label{56}
\eeq

\noindent Thus, via the consistency conditions we have arrived at the conclusion that not all configurations are in agreement with the perturbation. The consistency conditions serve as a kind of a selection rule for the zeroth order configurations 'matching' the specific perturbation (\ref{18}). The situation here is similar to what one encounters while considering perturbations of the quantum mechanical systems with a degenerate level, where the perturbation is well known to single out a basis in the subspace of the Hilbert space corresponding to that level. So, each of the pairs $(\Theta,\Psi)$ given above represents the zeroth order configuration singled out by the perturbation $\delta\mathcal{F}$. The first two configurations (\ref{51},\ref{52}) correspond to the wedge disclination lines, positive and negative, respectively, whereas the last four cases are twist configurations. 

Let us comment on the result for $\alpha=3$, see Eq. (\ref{49}). That identity suggests that the subgroup $SO(2)$ of the rotations about the direction of the disclination line may well represent the symmetry of the full system. Yet, one can check that, in general, those rotations do affect the value of the perturbation. So, all one can gather from Eq. (\ref{49}) is that they do not change the free energy up to the first order of the perturbative expansion. 

In conclusion, it is only some of the zeroth order solutions $\hat{q}_0^{(\Theta,\Psi)}$ that are 'matched' with the considered perturbation. In other words, the perturbative expansion of the order parameter will be consistent only when carried out around the selected configurations given by (\ref{51}-\ref{56}). In the next subsection we present a systematic procedure for finding the consecutive corrections to the order parameter.    

\subsection{Corrections to the order parameter}\label{opcor}

The corrections to the order parameter field in the consecutive orders of the perturbative expansion represent the solutions of the linear differential equations (\ref{27}), where the operator $\hat{\mathcal{L}}$, given by (\ref{28}), is the same for a fixed $\hat{q}_0$ in all orders of the expansion, and the inhomogeneous terms $\hat{\mathcal{N}}_k$ ($k=1,2,\ldots$) are defined in (\ref{29}), (\ref{30}). 

Let us begin with the construction of the solution for $\Theta=\Psi=0$ (a positive wedge disclination line), and $\Theta=\pi, \Psi=0$ (a negative wedge disclination line). Next we will pass on to discussing the remaining cases (\ref{53}-\ref{56}). It is convenient to introduce the local $\phi$-dependent basis in the space $\mathcal{M}$ defined by the following series of formulae

\be
\hat{E}_a^{\sigma}(\phi)=\mathcal{O}(\phi;\sigma)\hat{E}_{a}^{0}\mathcal{O}^{T}(\phi;\sigma),\label{57}
\ee

\noindent where $a=1,\ldots,5$, and 

\be
\mathcal{O}(\phi;\sigma)=\left(\begin{array}{ccc}
\cos\frac{\phi}{2} & -\sigma\sin\frac{\phi}{2} & 0 \\
\sigma\sin\frac{\phi}{2} & \cos\frac{\phi}{2} & 0 \\
0 & 0 & 1
\end{array}\right),\label{58}
\ee  
    
\noindent while

\be
\begin{array}{ccc}
\hat{E}_1^0=\frac{1}{\sqrt{6}}\left(\begin{array}{ccc}
1 & 0 & 0 \\
0 & 1 & 0 \\
0 & 0 & -2
\end{array}\right), &
\hat{E}_2^0=\frac{1}{\sqrt{2}}\left(\begin{array}{ccc}
1 & 0 & 0 \\
0 & -1 & 0 \\
0 & 0 & 0
\end{array}\right), &
\hat{E}_3^0=\frac{1}{\sqrt{2}}\left(\begin{array}{ccc}
0 & 1 & 0 \\
1 & 0 & 0 \\
0 & 0 & 0
\end{array}\right), \end{array} \label{59}
\ee

\be
\begin{array}{cc}
\hat{E}_4^0=\frac{1}{\sqrt{2}}\left(\begin{array}{ccc}
0 & 0 & 1 \\
0 & 0 & 0 \\
1 & 0 & 0 
\end{array}\right), &
\hat{E}_5^0=\frac{1}{\sqrt{2}}\left(\begin{array}{ccc}
0 & 0 & 0 \\
0 & 0 & 1 \\
0 & 1 & 0 
\end{array}\right)\end{array}.\label{60}
\ee 

\noindent The index $\sigma=\pm 1 $ corresponds to the case of the positive and negative wedge disclination line, respectively. The set $\{\hat{E}^{\sigma}_{a}(\phi), a=1,\ldots, 5\}$ is an orthonormal basis with respect to the standard scalar product in matrix spaces 

\be
\langle\hat{Q}_1|\hat{Q}_2\rangle=\mathrm{Tr}(\hat{Q}_1\hat{Q}_2).\label{61}
\ee
 
\noindent  We find it convenient to adopt here and in what follows the quantum mechanical bra-ket notation. 

Next step consists in projecting Eq. (\ref{27}) onto the directions of the basis vectors, which results in the following set of equation

\be
\langle\hat{E}_a^{\sigma}|\hat{\mathcal{L}}^{\sigma}\hat{q}_k^{\sigma}\rangle=\langle\hat{E}_a^{\sigma}|\mathcal{N}_k^{\sigma}\rangle,\label{62} 
\ee  

\noindent where $a=1,\ldots,5$, $k=1,2,\ldots $, and there is no summation with respect to $\sigma$.  Inserting between $\hat{\mathcal{L}}^{\sigma}$ and $\hat{q}_{k}$ the appropriate identity operator $|\hat{E}_b^{\sigma}\rangle\langle\hat{E}_b^{\sigma}|$, where the summation is carried out exclusivly over $b=1,\ldots,5$, one obtains a system of five equations 

\be
\mathcal{L}_{ab}^{\sigma}q_{kb}^{\sigma}=\mathcal{N}_{ka}^{\sigma}, \ \ \ \ \ \ \ a=1,\ldots, 5,\label{63} 
\ee

\noindent where 

\be
\mathcal{L}_{ab}^{\sigma}=
\langle\hat{E}_a^{\sigma}|\hat{\mathcal{L}}^{\sigma}|\hat{E}_b^{\sigma}\rangle=\mathrm{Tr}(\hat{E}_a^{\sigma}\hat{\mathcal{L}}^{\sigma}\hat{E}_b^{\sigma}),\label{64}
\ee

\be
\begin{array}{c}
q_{ka}^{\sigma}=\langle\hat{E}_a^{\sigma}|\hat{q}_k^{\sigma}\rangle=\mathrm{Tr}(\hat{q}_k^{\sigma}\hat{E}_a^{\sigma}), \\
\mathcal{N}_{ka}^{\sigma}=\langle\hat{E}_a^{\sigma}|\hat{\mathcal{N}}_k^{\sigma}\rangle=\mathrm{Tr}(\hat{\mathcal{N}}_k^{\sigma}\hat{E}_a^{\sigma}).\label{65}
\end{array}
\ee

\noindent It is clear, see (\ref{28}), that the matrix elements of $\hat{\mathcal{L}}^{\sigma}$ are not simple scalars but they contain differentiation with respect to polar coordinates $(s,\phi)$. Some careful algebra allows one to obtain 

\be
\mathcal{L}_{ab}^{\sigma}=\delta_{ab}\left(\Delta_{s}+\frac{1}{s^2}\frac{\partial^2}{\partial\phi^2}\right)-\frac{1}{s^2}\mathcal{J}_{1ab}+\sigma\frac{2}{s^2}\mathcal{J}_{2ab}\frac{\partial}{\partial\phi}+\mathcal{J}_{3ab},\label{66}
\ee
   
\noindent where $a,b=1,\ldots, 5$, $\delta_{ab}$ is the Kronecker delta, $\Delta_{s}=\frac{1}{s}\frac{\partial}{\partial s}\left(s\frac{\partial}{\partial s}\right)$, and matrices $\hat{\mathcal{J}}_1, \hat{\mathcal{J}}_2$ have the following form

\be
\begin{array}{cc}
\hat{\mathcal{J}}_1=\left(\begin{array}{ccccc}
0 & 0 & 0 & 0 & 0 \\
0 & 1 & 0 & 0 & 0 \\
0 & 0 & 1 & 0 & 0 \\
0 & 0 & 0 & \frac{1}{4} & 0 \\
0 & 0 & 0 & 0 & \frac{1}{4} 
\end{array}\right), & 
\hat{\mathcal{J}}_2=\left(\begin{array}{ccccc}
0 & 0 & 0 & 0 & 0 \\
0 & 0 & -1 & 0 & 0 \\
0 & 1 & 0 & 0 & 0 \\
0 & 0 & 0 & 0 & -\frac{1}{2} \\
0 & 0 & 0 & \frac{1}{2} & 0 
\end{array}\right)
\end{array}.\label{67}
\ee

\noindent Non-vanishing matrix elements of $\hat{\mathcal{J}}_{3}$ are given by the formulae

\be
\begin{array}{l}
\mathcal{J}_{311}=\frac{2}{3}-2\beta S-\frac{1}{2}(1+3\beta)(S^2+R^2), \\
\mathcal{J}_{312}=\mathcal{J}_{321}=\frac{1}{\sqrt{3}}R[6\beta-(1+3\beta)S], \\ \mathcal{J}_{322}=\frac{2}{3}+2\beta S-\frac{1}{6}(1+3\beta)(S^2+9R^2), \\
\mathcal{J}_{333}=\frac{2}{3}+2\beta S-\frac{1}{6}(1+3\beta)(S^2+3R^2), \\
\mathcal{J}_{344}=\frac{2}{3}-\beta(S-3R)-\frac{1}{6}(1+3\beta)(S^2+3R^2), \\
\mathcal{J}_{355}=\frac{2}{3}-\beta(S+3R)-\frac{1}{6}(1+3\beta)(S^2+3R^2),
\end{array}\label{68}
\ee
 
\noindent where parameter $\beta$ is defined in (\ref{10a}), and functions $S,R$ define the zeroth order structure of the order parameter, see (\ref{7}). 

Note that the only $\phi$-dependence of the matrix elements $\mathcal{L}_{ab}^{\sigma}$ in the basis $\hat{E}_{a}^{\sigma}$ lies in differential operators $\frac{\partial^2}{\partial\phi^2}$, $\frac{\partial}{\partial\phi}$. That leads to a great simplification in the next step, where the Fourier analysis of components $q_{ka}^{\sigma}(s,\phi)$ will be undertaken. Moreover, the formulae (\ref{66}-\ref{68}) show that the full space  $\mathcal{M}\ni\hat{q}^{\sigma}_{k}$ splits into the simple sum of two subspaces $\mathcal{M}=\mathcal{M}^{(1)}\oplus\mathcal{M}^{(2)}$ invariant with respect to the action of $\hat{\mathcal{L}}^{\sigma}$ therein. The subspace $\mathcal{M}^{(1)}$ comprises components $q_{ka}^{\sigma}$ with $a=1,2,3$, whereas $\mathcal{M}^{(2)}$ - the remaining ones. We will address the subspaces separately.

For each $k=1,2,\ldots$ we expand components $\{q_{ka}^{\sigma}, a=1,2,3\}$ of $\mathcal{M}^{(1)}$ in the Fourier series with respect to $\phi$

\be
q_{ka}^{\sigma}(s,\phi)=v_{ka}^{\sigma 0}(s)+\sum_{l=1}^{\infty}\left[v_{ka}^{\sigma l}(s)\cos(l\phi)+u_{ka}^{\sigma l}(s)\sin(l\phi)\right].\label{69} 
\ee

\noindent The Fourier analysis results in a fine structure of an infinite-dimensional space $\mathcal{F}^{(1)}$, which splits into an infinite simple sum of subspaces $\mathcal{F}_{l}^{\sigma(1)}$ invariant with respect to the action of $\hat{\mathcal{L}}^{\sigma{(1)}}$, where superscript ${(1)}$ refers to that part of $\hat{\mathcal{L}}^{\sigma}$ which acts in $\mathcal{M}^{(1)}$. One may write then

\be
\mathcal{F}^{(1)}=\bigoplus_{l=0}^{\infty}\mathcal{F}^{\sigma (1)}_{l}.\label{70}
\ee

\noindent The subspace $\mathcal{F}^{\sigma (1)}_{0}$ has the following structure

\be
\mathcal{F}^{\sigma (1)}_{0}=\left[\mathcal{S}\hat{E}^{\sigma}_1\right]\oplus\left[\mathcal{S}\hat{E}^{\sigma}_2\right]\oplus\left[\mathcal{S}\hat{E}^{\sigma}_3\right],\label{71}
\ee

\noindent where $\mathcal{S}$ denotes the space of smooth functions of the radial variable $s$, whereas the structure of the subspaces with $l\neq 0$ can be described as 

\beq
\mathcal{F}^{\sigma (1)}_{l}&=&\left[\mathcal{S}\hat{E}^{\sigma}_1\cos(l\phi)\right]\oplus\left[\mathcal{S}\hat{E}^{\sigma}_2\cos(l\phi)\right]\oplus\left[\mathcal{S}\hat{E}^{\sigma}_3\cos(l\phi)\right]\oplus \nonumber \\ 
&\oplus &\left[\mathcal{S}\hat{E}^{\sigma}_1\sin(l\phi)\right]\oplus\left[\mathcal{S}\hat{E}^{\sigma}_2\sin(l\phi)\right]\oplus\left[\mathcal{S}\hat{E}^{\sigma}_3\sin(l\phi)\right].\label{72}
\eeq

\noindent So, the three-dimensional subspace $\mathcal{F}_{0}^{\sigma(1)}$ comprises components  $\{v_{ka}^{\sigma 0}(s), a=1,2,3\}$, whereas the components belonging to the six-dimensional subspaces $\mathcal{F}^{\sigma (1)}_{l}$ ($l\neq 0$) are $\{v_{ka}^{\sigma l}(s),u_{ka}^{\sigma l}(s),\\ a=1,2,3\}$. The action of $\hat{\mathcal{L}}^{\sigma (1)}$ in $\mathcal{F}^{\sigma (1)}_{l}$ is represented by $\hat{\mathcal{L}}^{\sigma (1)}_{l}$, which for $l\neq 0$ are $6\times 6$ matrices with the following block structure

\be
\hat{\mathcal{L}}^{\sigma (1)}_{l}=\left(\begin{array}{c|c}
 \left(\Delta_s-\frac{l^2}{s^2}\right)I-\frac{1}{s^2}\hat{\mathcal{J}}_1^{(1)}+\hat{\mathcal{J}}_3^{(1)} & \sigma\frac{2 l}{s^2}\hat{\mathcal{J}}_2^{(1)} \\
\mbox{} \\  \hline \mbox{} \\
 -\sigma\frac{2 l}{s^2}\hat{\mathcal{J}}_2^{(1)} &  \left(\Delta_s-\frac{l^2}{s^2}\right)I-\frac{1}{s^2}\hat{\mathcal{J}}_1^{(1)}+\hat{\mathcal{J}}_3^{(1)}
\end{array}\right),\label{73} 
\ee

\noindent where $I$ in the formula above denotes a $3\times 3$ identity matrix, whereas $\hat{\mathcal{J}}^{(1)}_1$, $\hat{\mathcal{J}}^{(1)}_{2}$ and $\hat{\mathcal{J}}^{(1)}_{3}$ represent submatrices of $\hat{\mathcal{J}}_1$, $\hat{\mathcal{J}}_{2}$ and $\hat{\mathcal{J}}_{3}$, respectively, comprising the first three rows and columns. For $l=0$ we obtain the $3\times 3$ matrix given by the following formula

\be
\hat{\mathcal{L}}^{\sigma (1)}_{0}=\Delta_s I -\frac{1}{s^2}\hat{\mathcal{J}}^{(1)}_1+\hat{\mathcal{J}}^{(1)}_3.\label{74}
\ee

\noindent Note that this operator is independent of $\sigma$.

The last step is a partial diagonalization of $\hat{\mathcal{L}}^{\sigma (1)}_{l}$ with $l\neq 0$. It is carried out so as to get rid of the off-diagonal blocks containing $\hat{\mathcal{J}}^{(1)}_2$. Because $\hat{\mathcal{L}}^{\sigma (1)}_{l}$ are symmetric ($\left(\hat{\mathcal{J}}_2^{(1)}\right)^{\mathrm{T}}=-\hat{\mathcal{J}}^{(1)}_2$), the appropriate transformation $\mathcal{O}^{(1)}$  exists and, for both values of $\sigma$ and all nonzero $l$, has the same form 

\be
\mathcal{O}^{(1)}=\left(\begin{array}{cccccc}
1 & 0 & 0 & 0 & 0 & 0 \\
0 & \frac{1}{\sqrt{2}} & \frac{1}{\sqrt{2}} & 0 & 0 & 0 \\
0 & 0 & 0 & 0 & \frac{1}{\sqrt{2}} & -\frac{1}{\sqrt{2}} \\
0 & 0 & 0 & 1 & 0 & 0 \\
0 & 0 & 0 & 0 & \frac{1}{\sqrt{2}} & \frac{1}{\sqrt{2}} \\
0 & -\frac{1}{\sqrt{2}} & \frac{1}{\sqrt{2}} & 0 & 0 & 0 \\
\end{array}\right).\label{75}
\ee

\noindent On applying this transformation, $\hat{\mathcal{L}}^{\sigma (1)}_{l}$ take on the following form

\be
\hat{\mathcal{L}}^{\prime\sigma (1)}_{l}=(\mathcal{O}^{(1)})^{T}\hat{\mathcal{L}}^{\sigma (1)}_{l}\mathcal{O}^{(1)}=\left(\begin{array}{c|c}
\hat{\mathcal{D}}^{\sigma (1)}_{l}+\hat{\mathcal{J}}^{(1)} & \mathbf{0} \\
\mbox{ } & \mbox{ } \\
\hline
\mbox{ } & \mbox{ } \\
\mathbf{0} & \hat{\mathcal{D}}^{\sigma (1)}_{l}+\hat{\mathcal{J}}^{(1)} 
\end{array}\right),\label{76}
\ee

\noindent where $\mathbf{0}$ denotes the $3\times 3$ null matrix, and $\hat{\mathcal{D}}^{\sigma (1)}_{l}$ i $\hat{\mathcal{J}}^{(1)}$ are given by the formulae

\be
\hat{\mathcal{D}}^{\sigma (1)}_{l}=\left(\begin{array}{ccc}
\Delta_s-\frac{l^2}{s^2} & 0 & 0 \\
0 & \Delta_s-\frac{(1-\sigma l)^2}{s^2} & 0 \\
0 & 0 & \Delta_{s}-\frac{(1+\sigma l)^2}{s^2} 
\end{array}\right),\label{77}
\ee

\be
\hat{\mathcal{J}}^{(1)}=\left(\begin{array}{ccc}
j_1 & j_3 & j_3 \\
j_3 & j_2 & j_4 \\
j_3 & j_4 & j_2 
\end{array}\right),\label{78}
\ee 

\noindent where

\be
\begin{array}{l}
j_1=\frac{2}{3}-2\beta S-\frac{1}{2}(1+3\beta)(S^2+R^2), \\
j_2=\frac{2}{3}+2\beta S-\frac{1}{6}(1+3\beta)(S^2+6R^2), \\
j_3=\frac{1}{\sqrt{6}}\left[6\beta-(1+3\beta)S\right]R, \\
j_4=-\frac{1}{2}(1+3\beta)R^2
\end{array}.\label{79}
\ee

\noindent From (\ref{76}) one can see that each subspace $\mathcal{F}^{\sigma (1)}_{l}$ splits now into two invariant subspaces corresponding to the following pair of triplets of new components 

\be
\left\{\begin{array}{l}
v^{\prime\sigma l}_{k1}=v^{\sigma l}_{k1}, \\
v^{\prime\sigma l}_{k2}=\frac{1}{\sqrt{2}}\left(v^{\sigma l}_{k2}-u^{\sigma l}_{k3}\right), \\
v^{\prime\sigma l}_{k3}=\frac{1}{\sqrt{2}}\left(v^{\sigma l}_{k2}+u^{\sigma l}_{k3}\right), 
\end{array}\right.\label{80}
\ee

\noindent and 

\be
\left\{\begin{array}{l}
u^{\prime\sigma l}_{k1}=u^{\sigma l}_{k1}, \\
u^{\prime\sigma l}_{k2}=\frac{1}{\sqrt{2}}\left(v^{\sigma l}_{k3}+u^{\sigma l}_{k2}\right), \\
u^{\prime\sigma l}_{k3}=\frac{1}{\sqrt{2}}\left( -v^{\sigma l}_{k3}+u^{\sigma l}_{k2}\right).
\end{array}\right.\label{81}
\ee

\noindent Before making another step further, let us recapitulate the operations performed so far. Our aim is to solve the linear equation (\ref{63}) for consecutive corrections $\hat{q}_{k}^{\sigma}$ ($k=1,2,\ldots$) to the order parameter of the positive and negative ($\sigma=\pm 1$) wedge disclination line. The preliminary considerations have shown that the space $\mathcal{M}\ni\hat{q}_{k}^{\sigma}$ in the basis $\{\hat{E}_{a}^{\sigma}, a=1,\ldots , 5\}$ splits into the simple sum of two subspaces $\mathcal{M}^{(1)}$ i $\mathcal{M}^{(2)}$ invariant with respect to the action of $\hat{\mathcal{L}}^{\sigma}$ therein. The Fourier analysis of that part of $\hat{q}^{\sigma}_{k}$ which belongs to $\mathcal{M}^{(1)}$ has led to the structure of the space $\mathcal{F}^{(1)}=\bigoplus_{l=0}^{\infty}\mathcal{F}^{\sigma (1)}_{l}$.  Next, on carrying out the orthogonal transformation (\ref{75}) in each subspace $\mathcal{F}^{\sigma (1)}_{l}$ with $l\neq 0$, we arrive at the final form of the space $\mathcal{F}^{(1)}$ which now is the simple sum comprising the subspace $\mathcal{F}^{\sigma (1)}_{0}$, with $\hat{\mathcal{L}}^{\sigma (1)}_{0}$, given by (\ref{74}), representing the action of $\hat{\mathcal{L}}^{\sigma(1)}$ in this subspace, and two three-dimensional nonequivalent subspaces for each $l\neq 0$ where the action of $\hat{\mathcal{L}}^{\sigma(1)}$ is represented by $\hat{\mathcal{D}}^{\sigma (1)}_{l}+\hat{\mathcal{J}}^{(1)}$, see (\ref{77}), (\ref{78}). The next step consists in carrying out an analogous Fourier transformation for that part of $\hat{q}_{k}^{\sigma}$ which belongs to the subspace $\mathcal{M}^{(2)}$, \textit{i.e.} for components $\hat{q}^{\sigma}_{k4}$ and $\hat{q}^{\sigma}_{k5}$.  

Yet, there will be no full analogy in the Fourier expansion for components $\{\hat{q}^{\sigma}_{ka}, a=4,5\}$. On closer inspection of the r.h.s. terms in Eq. (\ref{63}), one can ascertain that the appropriate expansion is one comprising sines and cosines of the odd multiples of $\frac{\phi}{2}$, \textit{i.e.}

\be
q^{\sigma}_{ka}(s,\phi)=v^{\sigma 0}_{ka}(s)+\sum_{l=1}^{\infty}\left[v^{\sigma l}_{ka}(s)\cos\left(\frac{(2l-1)\phi}{2}\right)+u^{\sigma l}_{ka}(s)\sin\left(\frac{(2l-1)\phi}{2}\right)\right],\label{82}
\ee
 
\noindent Similarly, the module $\mathcal{F}^{(2)}$ generated by that expansion splits into the simple sum of the subspaces invariant with respect to the action of $\hat{\mathcal{L}}^{\sigma (1)}$ therein,

\be    
\mathcal{F}^{(2)}=\bigoplus_{l=0}^{\infty}\mathcal{F}^{\sigma (2)}_{l},\label{83}
\ee

\noindent which have the following structure

\be
\mathcal{F}^{\sigma (2)}_{0}=\left[\mathcal{S}\hat{E}^{\sigma}_{4}\right]\oplus\left[\mathcal{S}\hat{E}^{\sigma}_{5}\right],\label{84}
\ee

\noindent for $l=0$, and

\beq
\mathcal{F}^{\sigma (2)}_{l}&=&\left[\mathcal{S}\hat{E}^{\sigma}_{4}\cos\left(\frac{(2l-1)\phi}{2}\right)\right]\oplus\left[\mathcal{S}\hat{E}^{\sigma}_{5}\cos\left(\frac{(2l-1)\phi}{2}\right)\right]\oplus \nonumber \\ 
&\oplus &\left[\mathcal{S}\hat{E}^{\sigma}_{4}\sin\left(\frac{(2l-1)\phi}{2}\right)\right]\oplus\left[\mathcal{S}\hat{E}^{\sigma}_{5}\sin\left(\frac{(2l-1)\phi}{2}\right)\right],\label{85}
\eeq

\noindent for $l\neq 0$. So, the subspace $\mathcal{F}_{0}^{\sigma (2)}$ comprises components $\{v^{\sigma 0}_{ka}, a=4,5\}$, whereas in the four-dimensional subspace $\mathcal{F}^{\sigma (2)}_{l}$ the components $\{v^{\sigma l}_{ka}, u^{\sigma l}_{ka}, a=4,5\}$ are to be found. The action of $\hat{\mathcal{L}}^{\sigma (2)}$ in $\mathcal{F}^{\sigma (2)}_{l}$ is represented by $\hat{\mathcal{L}}^{\sigma (2)}_{l}$. For $l\neq 0$ the $4\times 4$ matrices of these operators have the following block structure

\be
\hat{\mathcal{L}}^{\sigma (2)}_{l}=\left(\begin{array}{c|c}
\left(\Delta_s-\frac{(2l-1)^2}{4 s^2}\right)I-\frac{1}{s^2}\hat{\mathcal{J}}^{(2)}_{1}+\hat{\mathcal{J}}^{(2)}_{3} & \sigma\frac{2l-1}{s^2}\hat{\mathcal{J}}^{(2)}_{2} \\
\mbox{} & \mbox{} \\
\hline
\mbox{} & \mbox{} \\
-\sigma\frac{2l-1}{s^2}\hat{\mathcal{J}}^{(2)}_{2} & \left(\Delta_s-\frac{(2l-1)^2}{4s^2}\right)I-\frac{1}{s^2}\hat{\mathcal{J}}^{(2)}_{1}+\hat{\mathcal{J}}^{(2)}_{3} \end{array}\right),\label{f86}
\ee

\noindent where $I$ in the above formula denotes a $2\times 2$ identity matrix, whereas $\hat{\mathcal{J}}^{(2)}_{1}$, $\hat{\mathcal{J}}^{(2)}_{2}$ and $\hat{\mathcal{J}}^{(2)}_{3}$ represent the submatrices of $\hat{\mathcal{J}}_1$, $\hat{\mathcal{J}}_{2}$ and $\hat{\mathcal{J}}_{3}$, respectively, comprising the last two rows and columns. The operator $\hat{\mathcal{L}}^{\sigma (2)}_{0}$ has the following form

\be 
\hat{\mathcal{L}}^{\sigma (2)}_{0}=\Delta_s I-\frac{1}{s^2}\hat{\mathcal{J}}^{(2)}_1+\hat{\mathcal{J}}^{(2)}_{3}.\label{87}
\ee
 
\noindent From (\ref{67}), (\ref{68}) one can see that this operator is diagonal and again independent of $\sigma$.

Likewise, the next step to take is the partial diagonalization of $\hat{\mathcal{L}}^{\sigma (2)}_{l}$ with $l\neq 0$. In each of the subspaces $\mathcal{F}^{\sigma (2)}_{l}$ we carry out such an orthogonal transformation $\mathcal{O}^{(2)}$ in order to get rid of the off-diagonal blocks in $\hat{\mathcal{L}}^{\sigma (2)}_{l}$. One can easily find that the right transformation $\mathcal{O}^{(2)}$ is identical for both values of $\sigma$ and all $l\neq 0$

\be
\mathcal{O}^{(2)}=\left(\begin{array}{cccc}
\frac{1}{\sqrt{2}} & \frac{1}{\sqrt{2}} & 0 & 0 \\
0 & 0 & \frac{1}{\sqrt{2}} & -\frac{1}{\sqrt{2}} \\
0 & 0 & \frac{1}{\sqrt{2}} & \frac{1}{\sqrt{2}} \\
-\frac{1}{\sqrt{2}} & \frac{1}{\sqrt{2}} & 0 & 0 
\end{array}\right).\label{88}
\ee

\noindent On applying this transformation to the operators $\hat{\mathcal{L}}^{\sigma (2)}_{l}$, they take on the form

\be
\hat{\mathcal{L}}^{\prime\sigma (2)}_{l}=(\mathcal{O}^{(2)})^{T}\hat{\mathcal{L}}^{\sigma (2)}_{l}\mathcal{O}^{(2)}=\left(\begin{array}{c|c}
\hat{\mathcal{D}}^{\sigma (2)}_{l}+\hat{\mathcal{J}}^{(2)} & \mathbf{0} \\
\mbox{} & \mbox{} \\
\hline
\mbox{} & \mbox{} \\
\mathbf{0} & \hat{\mathcal{D}}^{\sigma (2)}_{l}+\hat{\mathcal{J}}^{(2)} 
\end{array}\right),\label{89}
\ee

\noindent where $\mathbf{0}$ denotes the $2\times 2$ null matrix, whereas $\hat{\mathcal{D}}^{\sigma (2)}_{l}$ and $\hat{\mathcal{J}}^{(2)}$ are defined by the following formulae

\be
\hat{\mathcal{D}}^{\sigma (2)}_{l}=\left(\begin{array}{cc}
\Delta_s-\frac{[1-\sigma(2l-1)]^2}{4s^2} & 0 \\
0 & \Delta_s-\frac{[1+\sigma(2l-1)]^2}{4s^2} 
\end{array}\right),\label{90}
\ee

\be
\hat{\mathcal{J}}^{(2)}=\left(\begin{array}{cc}
\frac{2}{3}-\beta S -\frac{1}{6}(1+3\beta)(S^2+3R^2) & 3\beta R \\
3\beta R & \frac{2}{3}-\beta S -\frac{1}{6}(1+3\beta)(S^2+3R^2) 
\end{array}\right).\label{91}
\ee

\noindent From (\ref{89}) one can clearly see that each subspace $\mathcal{F}^{\sigma (2)}_{l}$ splits into two invariant subspaces corresponding to the following pairs of new components

\be
\left\{\begin{array}{l}
v^{\prime\sigma l}_{k4}=\frac{1}{\sqrt{2}}\left(v^{\sigma l}_{k4}-u^{\sigma l}_{k5}\right), \\
v^{\prime\sigma l}_{k5}=\frac{1}{\sqrt{2}}\left(v^{\sigma l}_{k4}+u^{\sigma l}_{k5}\right), 
\end{array}\right.\label{92}
\ee

\noindent and

\be
\left\{\begin{array}{l}
u^{\prime\sigma l}_{k4}=\frac{1}{\sqrt{2}}\left(v^{\sigma l}_{k5}+u^{\sigma l}_{k4}\right), \\
u^{\prime\sigma l}_{k5}=\frac{1}{\sqrt{2}}\left(-v^{\sigma l}_{k5}+u^{\sigma l}_{k4}\right).
\end{array}\right.\label{93}
\ee

This ends the first two stages of the construction, both of which were focused exclusively on the l.h.s. of Eqs. (\ref{63}). The last natural stage is, therefore, to transfer all the transformations performed to the r.h.s. of those equations. The right order of carrying out all these transformations and the transformations themselves follow clearly from the context of the analysis presented above. So, the only task remaining is to introduce appropriate and, we hope, clear notation. Vector $[\mathcal{N}^{\sigma}_{ka}]_{\{a=1,2,3\}}$ belongs to $\mathcal{M}^{(1)}$, whereas $[\mathcal{N}^{\sigma}_{ka}]_{\{a=4,5\}}$ -- to $\mathcal{M}^{(2)}$. The Fourier analysis in these spaces leads to the spaces $\mathcal{F}^{(1)}$ and $\mathcal{F}^{(2)}$ where the components $V^{\sigma l}_{ka}(s)$ i $U^{\sigma l}_{ka}(s)$ ($a=1,2,\ldots,5$) of the above vectors belong. They are given by the following formulae (see (\ref{71}), (\ref{72}), (\ref{84}), (\ref{85}))

\be
\begin{array}{l}
V^{\sigma 0}_{ka}(s)=\frac{1}{2\pi}\int_{0}^{2\pi}\mathcal{N}^{\sigma}_{ka}(s,\phi)d\phi, \\
V^{\sigma l}_{ka}(s)=\frac{1}{\pi}\int_{0}^{2\pi}\mathcal{N}^{\sigma}_{ka}(s,\phi)\cos (l\phi)d\phi, \\
U^{\sigma l}_{ka}(s)=\frac{1}{\pi}\int_{0}^{2\pi}\mathcal{N}^{\sigma}_{ka}(s,\phi)\sin (l\phi)d\phi
\end{array}\label{94}
\ee

\noindent for $a=1,2,3$, and

\be
\begin{array}{l}
V^{\sigma 0}_{ka}(s)=\frac{1}{2\pi}\int_{0}^{2\pi}\mathcal{N}^{\sigma}_{ka}(s,\phi)d\phi, \\
V^{\sigma l}_{ka}(s)=\frac{1}{\pi}\int_{0}^{2\pi}\mathcal{N}^{\sigma}_{ka}(s,\phi)\cos\left(\frac{(2l-1)\phi}{2}\right)d\phi, \\
U^{\sigma l}_{ka}(s)=\frac{1}{\pi}\int_{0}^{2\pi}\mathcal{N}^{\sigma}_{ka}(s,\phi)\sin\left(\frac{(2l-1)\phi}{2}\right)d\phi
\end{array}\label{95}
\ee

\noindent for $a=4,5$, and the following inclusions hold

\be
\begin{array}{l}
\vec{N}^{\sigma 0}_{k(1)}\equiv [V^{\sigma 0}_{ka}]_{\{a=1,2,3\}}\in\mathcal{F}^{\sigma (1)}_{0}, \\
\vec{N}^{\sigma l}_{k(1)}\equiv [V^{\sigma l}_{ka},U^{\sigma l}_{ka}]_{\{a=1,2,3\}}\in\mathcal{F}^{\sigma (1)}_{l}, \\
\vec{N}^{\sigma 0}_{k(2)}\equiv [V^{\sigma 0}_{ka}]_{\{a=4,5\}}\in\mathcal{F}^{\sigma (2)}_{0}, \\  
\vec{N}^{\sigma l}_{k(2)}\equiv [V^{\sigma l}_{ka}, U^{\sigma l}_{ka}]_{\{a=4,5\}}\in\mathcal{F}^{\sigma (2)}_{l}.
\end{array}\label{96}
\ee

\noindent The last step consists in taking the orthogonal transformations $\mathcal{O}^{(1)}$ and $\mathcal{O}^{(2)}$ ($l\neq 0$) into account. It is clear that the right transformation rules for $\vec{N}^{\sigma l}_{k (1)}$ and $\vec{N}^{\sigma l}_{k(2)}$ have the form

\be
\begin{array}{l}
\vec{N}^{\prime\sigma l}_{k(1)}=(\mathcal{O}^{(1)})^{T}\vec{N}^{\sigma l}_{k(1)}, \\
\vec{N}^{\prime\sigma l}_{k(2)}=(\mathcal{O}^{(2)})^{T}\vec{N}^{\sigma l}_{k(2)}.
\end{array}\label{97}
\ee  
 
In conclusion, the solution of the original Eq. (\ref{63}) in each order $k=1,2,\ldots$ of the perturbative expansion has been reduced to the solution of the following, in general, infinite set of ordinary differential equations

\be
\left\{\begin{array}{l}
\hat{\mathcal{L}}^{\sigma (1)}_{0}\vec{q}^{\ \sigma 0}_{k(1)}=\vec{N}^{\sigma 0}_{k(1)}, \\
\hat{\mathcal{L}}^{\prime\sigma (1)}_{l}\vec{q}^{\ \prime\sigma l}_{k(1)}=\vec{N}^{\prime\sigma l}_{k(1)}\ \ \  \ \mathrm{dla}\ \ \ l\neq 0, \\ 
\hat{\mathcal{L}}^{\sigma (2)}_{0}\vec{q}^{\ \sigma 0}_{k(2)}=\vec{N}^{\sigma 0}_{k(2)}, \\
\hat{\mathcal{L}}^{\prime\sigma (2)}_{l}\vec{q}^{\ \prime\sigma l}_{k(2)}=\vec{N}^{\prime\sigma l}_{k(2)}\ \ \ \ \mathrm{dla}\ \ \ l\neq 0,
\end{array}\right.\label{98}
\ee

\noindent where

\be
\begin{array}{l}
\vec{q}^{\ \sigma 0}_{k(1)}\equiv [v^{\sigma 0}_{ka}]_{\{a=1,2,3\}}, \\
\vec{q}^{\ \prime\sigma l}_{k(1)}\equiv [v^{\prime\sigma l}_{ka},u^{\prime\sigma l}_{ka}]_{\{a=1,2,3\}}\ \ \ \ \mathrm{dla}\ \ \ l\neq 0, \\
\vec{q}^{\ \sigma 0}_{k(2)}\equiv [v^{\sigma 0}_{ka}]_{\{a=4,5\}}, \\
\vec{q}^{\ \prime\sigma l}_{ka}\equiv [v^{\prime\sigma l}_{ka},u^{\prime\sigma l}_{ka}]_{\{a=4,5\}}\ \ \ \ \ \ \mathrm{dla}\ \ \ l\neq 0, 
\end{array}\label{99}
\ee

\noindent Fortunately, only a finite number of the inhomogeneous terms in (\ref{98}) are non-zero. For the detailed discussion of the first order corrections to the order parameter for both the positive and negative wedge disclination lines see Appendix \ref{ap1}. 

Now the construction of the solutions for the remaining four cases (\ref{53}-\ref{56})) representing the twist disclination lines is yet to be discussed. Let $\hat{q}^{\perp}$ denote the solution corresponding to these cases. In the zeroth order of the perturbative expansion the order parameter can be written as follows

\be 
\hat{q}^{\perp}_{0}=\mathcal{O}_1(\Theta,\Psi)\hat{q}^{+}_{0}\left(\mathcal{O}_1(\Theta,\Psi)\right)^{T},\label{100}
\ee

\noindent where $(\Theta, \Psi)$ is one of the following pairs

\be
\left\{\left(\frac{\pi}{2},0\right),\left(\frac{\pi}{2},\frac{\pi}{4}\right), \left(\frac{\pi}{2},\frac{\pi}{2}\right),\left(\frac{\pi}{2},\frac{3\pi}{4}\right)\right\}.\label{101}
\ee

\noindent So, the relevant equations for the consecutive corrections to the order parameter $\hat{q}_{k}^{\perp}$ have the form

\be
\begin{array}{c}
\tilde{\partial}_{k}\tilde{\partial}_{k}\hat{q}^{\perp}_{k}+\frac{2}{3}\hat{q}^{\perp}_{k}+3\beta\left(\hat{q}^{\perp}_{0}\hat{q}^{\perp}_{k}+\hat{q}^{\perp}_{k}\hat{q}^{\perp}_{0}\right)-\frac{1}{2}(1+3\beta)\mathrm{Tr}\left(\hat{q}^{\perp}_{0}\hat{q}^{\perp}_{k}\right)\hat{q}^{\perp}_{0}- \\ 
-\frac{1}{4}(1+3\beta)\mathrm{Tr}\left(\hat{q}^{\perp}_{0}\right)^2\hat{q}^{\perp}_{k}-2\beta\mathrm{Tr}\left(\hat{q}^{\perp}_{0}\hat{q}^{\perp}_{k}\right)I=\hat{\mathcal{N}}^{\perp}_{k}.
\end{array}\label{102}
\ee

\noindent Let us multiply either side of the above equation by $\left(\mathcal{O}_1(\Theta,\Psi)\right)^{T}$ and $\mathcal{O}_{1}(\Theta,\Psi)$ from the left and right side, respectively. Then, one obtains

\be
\begin{array}{c}
\tilde{\partial}_{k}\tilde{\partial}_{k}\hat{\tilde{q}}^{\perp}_{k}+\frac{2}{3}\hat{\tilde{q}}^{\perp}_{k}+3\beta\left(\hat{q}^{+}_{0}\hat{\tilde{q}}^{\perp}_{k}+\hat{\tilde{q}}^{\perp}_{k}\hat{q}^{0}_{0}\right)-\frac{1}{2}(1+3\beta)\mathrm{Tr}\left(\hat{q}^{+}_{0}\hat{\tilde{q}}^{\perp}_{k}\right)\hat{q}^{+}_{0}- \\ 
-\frac{1}{4}(1+3\beta)\mathrm{Tr}\left(\hat{q}^{+}_{0}\right)^2\hat{\tilde{q}}^{\perp}_{k}-2\beta\mathrm{Tr}\left(\hat{q}^{+}_{0}\hat{\tilde{q}}^{\perp}_{k}\right)I=\hat{\tilde{\mathcal{N}}}^{\perp}_{k},
\end{array}\label{103}
\ee

\noindent where

\beq
\hat{\tilde{q}}^{\perp}_{k}&=&\left(\mathcal{O}_{1}(\Theta,\Psi)\right)^{T}\hat{q}^{\perp}_{k}\mathcal{O}_{1}(\Theta,\Psi),\label{104} \\
\hat{\tilde{\mathcal{N}}}^{\perp}_{k}&=&\left(\mathcal{O}_{1}(\Theta,\Psi)\right)^{T}\hat{\mathcal{N}}^{\perp}_{k}\mathcal{O}_{1}(\Theta,\Psi),\label{105} 
\eeq
 
\noindent or in symbolic notation

\be
\hat{\mathcal{L}}^{+}\hat{\tilde{q}}^{\perp}_k=\hat{\tilde{\mathcal{N}}}^{\perp}_{k}.\label{106}
\ee

\noindent As a result, we have reduced the problem of solving Eqs. (\ref{102}) for $\hat{q}_k^{\perp}$ to the previously discussed problem of solving Eqs. (\ref{106}) for the modified corrections $\hat{\tilde{q}}^{\perp}_{k}$, with exactly the same linear operator $\hat{\mathcal{L}}^{+}$ as for the positive wedge disclination line. So, the main part of the analysis will proceed as it did for the case of $\sigma=+1$, the only difference lying in the appropriately adjusted notation; all quantities corresponding to the corrections to the order parameter should be fitted with a tilde and superscribed with $\perp$ instead of $\sigma$. For the first order correction to the order parameter of the twist disclination line with $(\Theta,\Psi)=(\frac{\pi}{2},0)$ see Appendix \ref{ap2}. 

The general solution for the corrections $\hat{q}_{k}$ ($k=1,2,\ldots$) in each of the cases discussed above consists of two terms, \textit{i.e.} of a particular solution of the relevant inhomogeneous equation and a general solution of its auxiliary homogeneous partner, the latter spanning an infinite-dimensional space. That space splits into the subspaces corresponding each to the fixed value of $l=0,2,\ldots$ and one of the indices $(1)$ or $(2)$. Obviously, only those elements which are continuous at $s=0$ and finite for $s\rightarrow\infty$ are physically relevant and should be incorporated in the solution. In general, they correspond to different forms of excitations of the disclination line, the zero modes $\hat{\tau}_{i}$ ($i=1,2,3$), $\hat{\theta}_{\alpha}$ ($\alpha=1,2,3$) among them enjoying a particularly simple physical interpretation. In the first order we set the contribution from these solutions to zero. One can check, not easily though, that this is consistent with the constraints (\ref{41}) for $k=2$. So, the final solution for the rectilinear disclination line up to the first order of the expansion has the form

\be
\hat{q}^{\Sigma}=\hat{q}^{\Sigma}_0+\varepsilon\hat{q}^{\Sigma}_1,\label{107}
\ee

\noindent where $\Sigma\in\{+, -, \star\}$ is to refer to the different types of the disclination line, see Appendix. The solution (\ref{107}) may be called the minimal solution in the sense that it contains no excitations except for those that are forced by the anisotropic perturbation. 

For the sake of completness of the investigation, we have found numerical solutions for the first order corrections to the order parameter of the wedge disclination lines in the special high symmetry case of $\beta=\frac{1}{3}$. Then, $S\equiv 1$ and the only non-vanishing functions $P^{\pm}_{3}(s)$ (see Appendix \ref{ap1}) satisfy the following equations

\beq
& &P_3^{+\prime\prime}+\frac{1}{s}P_3^{+\prime}-\frac{1}{s^2}P_3^{+}-(1+R^2)P_3^{+}=-\frac{1}{6}\left(R^{\prime\prime}+\frac{1}{s}R^{\prime}-\frac{1}{s^2}R\right),\label{108}\\
& & \nonumber \\
& &P_3^{-\prime\prime}+\frac{1}{s}P_3^{-\prime}-\frac{9}{s^2}P_3^{-}-(1+R^2)P_3^{-}=-\frac{1}{6}\left(R^{\prime\prime}-\frac{3}{s}R^{\prime}+\frac{3}{s^2}R\right)\label{109}.
\eeq

\noindent The numerical solutions were found with the help of the \textit{Mathematica 3.0} package. Below, the zeroth order structure function $R(s)$ as well as the first order structure functions $P^{\pm}_{3}(s)$ together with their first derivatives are depicted.

\begin{figure}[h!]
\centerline{\psfig{figure=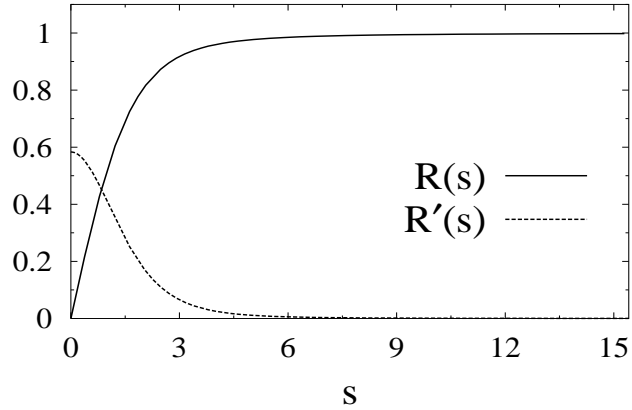,width=10 cm,height= 6.5 cm}}
\caption{Functions $R(s)$ and $R'(s)$. The numerical solution has been generated within the interval $(0.0001,18)$; for $s\in (0,0.0001)$ the numerical solution has been appended with the polynomial approximation smoothly matched at $s=0.0001$.}
\label{graph1}
\end{figure}

\begin{figure}[h!]
\centerline{\psfig{figure=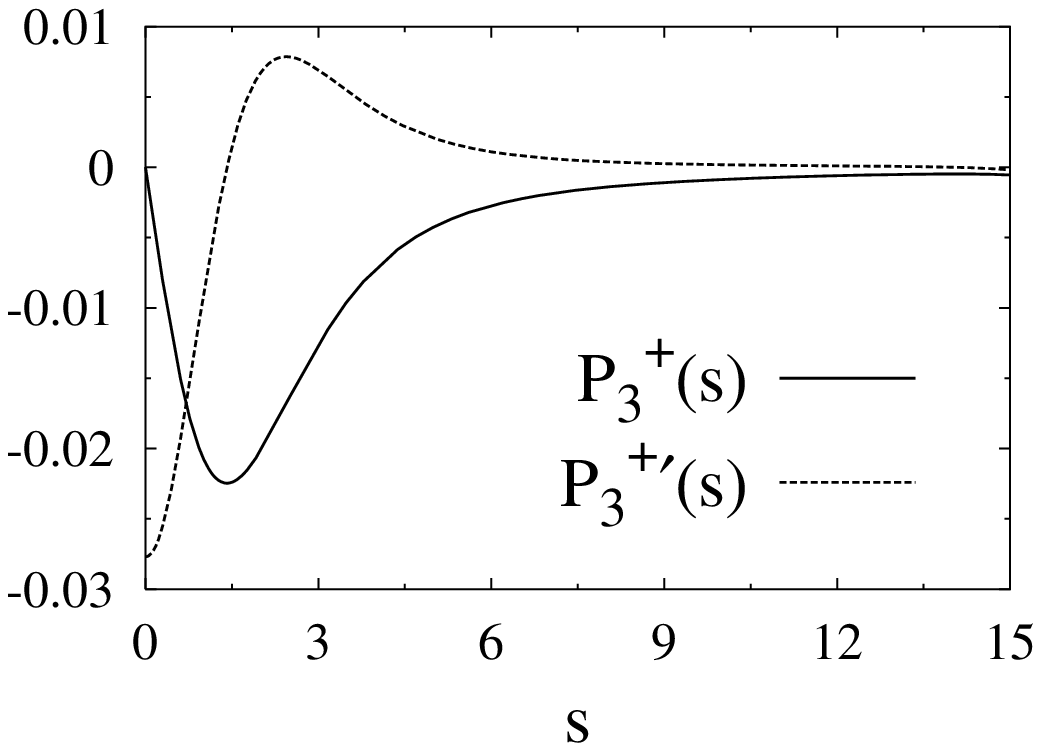,width=10.5 cm,height=6.5 cm}}
\caption{Functions $P_3^{+}(s)$ and $P_3^{+\prime}(s)$. The numerical solution has been found in the interval $(0.2,15)$; for $S\in (0,0.2)$ the numerical solution has been appended with the polynomial approximation smoothly matched at $s=0.2$.}
\label{graph2}
\end{figure}

\begin{figure}[h!]
\centerline{\psfig{figure=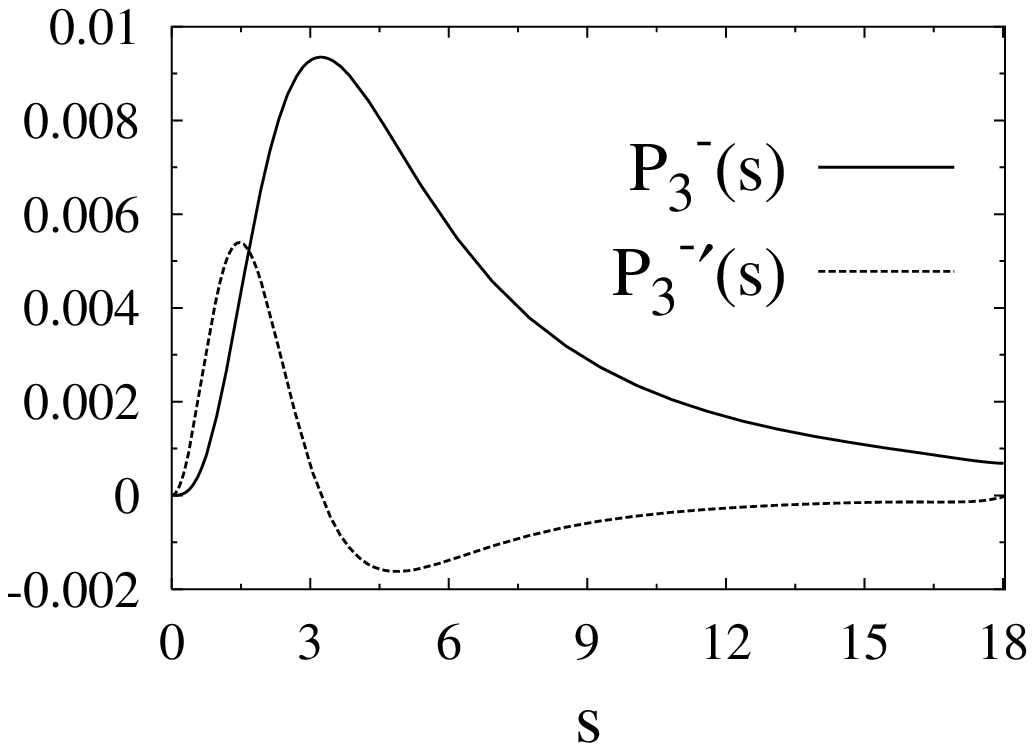,width=11 cm,height=6.5 cm}}
\caption{Functions $P_3^{-}(s)$ and $P_3^{-\prime}(s)$. The numerical solution has been generated in the interval $(0.5,18)$; for $s\in (0,0.5)$ the numerical solution has been appended with the polynomial approximation smoothly matched at $s=0.5$.}
\label{graph3}
\end{figure}

\noindent The graphs above show that the first order corrections are of order $\sim 0.01\varepsilon$. Taking into account the estimated values of the perturbative parameter (see the beginnig of this section) and the order of magnitude of the unperturbed contribution, one can see that the ratio of the first to zeroth order contribution is $\sim 0.01$. Therefore, the idea of the perturbative scheme finds quite a good justification. Admittedly, the above conclusion  holds for the special case of $\beta=\frac{1}{3}$, but the equations for the functions $P(s)$ giving the first order corrections to the order parameter (see Appendix) depend smoothly on $\beta$ and, therefore,  one can expect this conclusion to hold even in the generic case when $\beta\neq\frac{1}{3}$. 
 
\section{Corrections to the free energy}\label{fecor} 

In general, the knowledge of the corrections to the order parameter up to the $k$-th order of the expansion in the discussed model allows one to calculate the $(k+1)$-th order correction to the free energy. The free energy of the smooth disclination line contained in the concentric cylinder of the radius $L\tilde{\xi}_0$ per unit length is given by the formula

\be
F_b=2\pi\tilde{\xi}_0^2\ \frac{9}{8}a\eta_0^2\ \kappa_b,\label{110}
\ee
  
\noindent where

\be
\begin{array}{c}
\kappa_b=\frac{1}{2\pi}\int_{\mathbf{K}(0,L)}d^2\vec{s}\left\{\frac{3}{2}\left[\tilde{\partial}_{k}q_{ij}\tilde{\partial}_{k}q_{ij}+\varepsilon\left(\tilde{\partial}_{i}q_{ik}\tilde{\partial}_{j}q_{jk}-\frac{1}{2}\tilde{\partial}_{k}q_{ij}\tilde{\partial}_{k}q_{ij}\right)\right]\right. \\
\mbox{} \\
\left. -\left[\mathrm{Tr}\left(\hat{q}^2\right)+3\beta\mathrm{Tr}\left(\hat{q}^3\right)-\frac{3}{16}(1+3\beta)\left(\mathrm{Tr}\left(\hat{q}^2\right)\right)^2\right]\right\}
\end{array}\label{111}
\ee

\noindent is a dimensionless quantity, and the subscript 'b' is meant to refer to the smoothness of the underlying configuration. Inserting the expansion (\ref{23}) into (\ref{111}), we obtain

\be
\kappa_b=\kappa_{b0}+\varepsilon\kappa_{b1}+\varepsilon^2\kappa_{b2}+\ldots,\label{112}
\ee

\noindent where

\be
\kappa_{bk}=\frac{1}{2\pi}\int_{\mathbf{C}(0,L)}d^2\vec{s}f_{bk}\label{113}
\ee

\noindent for $k=1,2,\ldots$; $f_{bk}$ represent the dimensionless corrections to the free energy density in the consecutive orders of the expansion. In the first two orders they have the form

\beq
f_{b0}&=&\frac{3}{2}\tilde{\partial}_{k}q_{0ij}\tilde{\partial}_{k}q_{0ij}-\left[\mathrm{Tr}\left(\hat{q}_0^2\right)+3\beta\mathrm{Tr}\left(\hat{q}_0^3\right)-\frac{3}{16}(1+3\beta)\left(\mathrm{Tr}\left(\hat{q}_0^2\right)\right)^2\right],\label{114} \\
f_{b1}&=&\frac{3}{2}\left[2\tilde{\partial}_{k}q_{0ij}\tilde{\partial}_{k}q_{1ij}+\tilde{\partial}_{i}q_{0ik}\tilde{\partial}_{j}q_{0jk}-\frac{1}{2}\tilde{\partial}_{k}q_{0ij}\tilde{\partial}_{k}q_{0ij}\right]- \nonumber \\
&-&\left[2\mathrm{Tr}\left(\hat{q}_0\hat{q}_1\right)+9\beta\mathrm{Tr}\left(\hat{q}_0^2\hat{q}_1\right)-\frac{3}{4}(1+3\beta)\mathrm{Tr}\left(\hat{q}_0^2\right)\mathrm{Tr}\left(\hat{q}_0\hat{q}_1\right)\right].\label{115} 
\eeq

The zeroth order contribution had been thoroughly discussed in \cite{disc1}. Let us quote here the most relevant results from \cite{disc1}. The zeroth order contribution to $\kappa_{b}$ can be written in the form

\be
\kappa_{b0}=\int_{0}^{s_0}ds sf_{b0}(s)+\int_{s_0}^{L}ds s f_{b0}^{\mathrm{reg}}(s)+\int_{s_0}^{L}ds s\left[\frac{3}{s^2}-\frac{4}{3}(1+\beta)\right],\label{117}
\ee 

\noindent where

\be
f_{b0}(s)=\frac{8}{9a\eta_0^2}\mathcal{F},\label{118}
\ee

\noindent where $\mathcal{F}$ has the following form

\beq
\mathcal{F}&=&\frac{9}{8}a\eta_0^2\left[{S'}^2+3{R'}^2+\frac{3}{s^2}R^2-\frac{2}{3}S^2\right. \nonumber \\
&-&\left.2R^2-2\beta S\left(3R^2-\frac{1}{3}S^2\right)+\frac{1}{12}(1+3\beta)(S^2+3R^2)^2\right].\label{119}
\eeq

\noindent The regularized free energy density is given by the formula

\be
\begin{array}{l}
f_{b0}^{\mathrm{reg}}(s)=\left(S^{\prime}\right)^2+3\left(R^{\prime}\right)^2-3\frac{(1-R)(1+R)}{s^2}+\frac{3}{4}(1+3\beta)(1-R)^2(1+R)^2+ \\
\hspace{1.3 cm} +\ \frac{1}{12}(1-S)^2\left[32\beta+8\beta(1+S)+(1+3\beta)(1+S)^2\right]-\label{120} \\
\hspace{1.3 cm} -\ \frac{1}{2}(1-S)(1-R)(1+R)\left[9\beta-1-(1+3\beta)S\right].
\end{array}
\ee

\noindent The choice of $s_0\in (0,L)$ in (\ref{117}) is completely arbitrary because it does not affect the value of $\kappa_{b0}$. It is apparent from (\ref{117}) that the free energy of the unperturbed disclination line comprises two divergent terms, see the third integral in (\ref{117}). The first one is logarithmically divergent and its appearance is due to the existence of the rotational zero modes. The other divergent term originates from the fact that the free energy density of the ordered phase takes a negative value relative to that of the isotropic phase which has been set to zero. 

The asymptotic behaviour of the free energy density of the singular disclination line in the zeroth order has the following form

\be
f_{c0}(s)=\frac{3}{s^2}-\frac{4}{3}(1+\beta),\label{121}
\ee

\noindent where the subscript 'c' had been meant to refer to singular configurations. Hence, the corresponding zeroth order contribution to $\kappa_c$ is given by the formula

\be
\kappa_{c0}=\int_{s_{c0}}^{L}ds sf_{c0}(s),\label{122}
\ee

\noindent where $s_{c0}$, representing the dimensionless radius of the singular core in the zeroth order of the perturbative expansion, has the form

\be
s_{c0}=\frac{3}{2\sqrt{1+\beta}}.\label{123}
\ee

The form of the first order correction to the free energy can be readily obtained by making use of the following identity

\beq
\int_{\mathbf{C}(0,L)}d^2\vec{s}\tilde{\partial}_{l}q_{0ij}\tilde{\partial}_{l}q_{kij}\!\!&=&\!\!\!\int_{\mathbf{C}(0,L)}d^2\vec{s}\left[\frac{2}{3}\mathrm{Tr}\left(\hat{q}_0\hat{q}_k\right)+3\beta\mathrm{Tr}\left(\hat{q}_0^2\hat{q}_k\right)-\frac{1}{4}(1+3\beta)\mathrm{Tr}\left(\hat{q}_0^2\right)\mathrm{Tr}\left(\hat{q}_0\hat{q}_k\right)\right] \nonumber \\
&+&\!\!\!\int_{\mathbf{\partial C}(0,L)}\mathrm{Tr}\left(\hat{q}_k\tilde{\partial}_l\hat{q}_0\right)d\tilde{\sigma}^l,\label{124}
\eeq   

\noindent which can be derived from Eq(\ref{24}) by first multiplying either side of it by $\hat{q}_k$, then taking the trace and integrating over the area of a circle $\mathbf{C}(0,L)$. The first order correction can then be written as

\be
2\pi\kappa_{b1}=\int_{\mathbf{C}(0,L)}d^2\vec{s}\tilde{f}_{b1}(\vec{s})+3\int_{\partial\mathbf{C}(0,L)}\mathrm{Tr}\left(\hat{q}_1\tilde{\partial}_{l}\hat{q}_0\right)d\tilde{\sigma}^{l},\label{125}
\ee   

\noindent where 

\be
\tilde{f}_{b1}(\vec{s})=\frac{3}{2}\left[\tilde{\partial}_{i}q_{0il}\tilde{\partial}_{j}q_{0jl}-\frac{1}{2}\tilde{\partial}_{l}q_{0ij}\tilde{\partial}_{l}q_{0ij}\right].\label{126}
\ee

\noindent The formula (\ref{125}) shows that to obtain $\kappa_{b1}$ one does not need to know the first order correction to the order parameter in detail. The only relevant information concerns the asymptotic behaviour of $\hat{q}_1$; for $\hat{q}_1^{\pm}$ and $\hat{q}_1^{\star}$ discussed in Appendix \ref{ap1} and \ref{ap2}, respectively, the surface contribution in (\ref{125}) vanishes at least as $L^{-2}$ in the limit $L\rightarrow\infty$. Now, taking advantage of the polynomial approximation for the functions $R$,$S$ developed in \cite{disc1}, we can estimate the first order corrections to the free energy of the smooth disclination lines discussed in Appendices. Using (\ref{125}), (\ref{126}), and the results from \cite{disc1}, one obtains 

\beq
\kappa_{b1}^{\sigma}&=& -\frac{1}{6}(w_0-1)^2+\sigma\frac{3}{2}+\mathcal{O}(L^{-2}),\label{127} \\
\kappa_{b1}^{\star}&=&-\frac{3}{16}-\frac{1}{24}(w_0-1)^2-\frac{3}{4}\ln\left(\frac{L}{s_1}\right)+\mathcal{O}(L^{-2}).\label{128}
\eeq

\noindent Firstly, from (\ref{127},\ref{128}) it follows clearly that the perturbation removes the degeneracy. Up to the first order the free energy gap between the positive and negative wedge disclination lines is given by

\be
\kappa_{b1}^{+}-\kappa_{b1}^{-}=3+\mathcal{O}(L^{-2})>0.\label{129}
\ee

\noindent But it is the twist disclination line that represents the configuration of the lowest free energy. This is due to the logarithmic divergence of the corresponding free energy density. That conclusion is in perfect agreement with the result obtained in \cite{anisimov}, where the corresponding solutions within the framework of the Oseen-Z\"{o}cher-Frank theory had been considered. 

The first order corrections to the free energy of the corresponding singular disclination lines one can find remembering that in that case the only 'active' degrees of freedom are those connected with the average orientation of nematogenic molecules, \textit{i.e.} the director field, whereas the degree of the orientational ordering itself is fixed at the vacuum value. Formally, the transition from the smooth configuration to the singular one is equivalent to taking the limit $R,S\rightarrow 1$. For both the wedge disclination lines this leads immediately to the following result  

\be
\kappa^{\sigma}_{c1}=\sigma\frac{3}{2}+\mathcal{O}(L^{-2}),\label{130}
\ee

\noindent which gives

\be
\Delta\kappa_{bc1}^{\sigma}(\beta)\equiv\kappa_{b1}^{\sigma}-\kappa_{c1}^{\sigma}=-\frac{1}{6}[w_0(\beta)-1]^2 +\mathcal{O}(L^{-2})\label{131}
\ee

\noindent Due to the presence of the logarithmically divergent term in (\ref{128}), the limit $R,S\rightarrow 1$ cannot be carried out in such a straightforward manner. In that case, one has to perform the transition at the level of the corresponding free energy density. Some strenuous algebra gives   

\be
f_{c1}^{\star}(s,\phi)= -\frac{3}{4}\frac{1}{s^2}\left(1+\cos 2\phi\right).\label{132}
\ee

\noindent for the first correction to the free energy density of the singular twist disclination line. Like in \cite{disc1}, we assume that the boundary of the core of the singular disclination line represents the set of points at which the free energy density vanishes, \textit{i.e} it is equal to the free energy density of the isotropic phase. So, the radius of the boundary as the function of $\phi$ is given by

\be
f_{c}^{\star}(s_c^{\star}(\phi),\phi)=0,\label{133}
\ee  

\noindent where $f_{c}^{\star}(s,\phi)$ is the full free energy density of the twist disclination line. A simple calculation gives

\be
s_{c}^{\star}(\phi)=s_{c0}-\frac{1}{8}\varepsilon s_{c0}(1+\cos 2\phi)+\mathcal{O}(\varepsilon^2).\label{134}
\ee

\noindent The first order correction $\kappa_{c1}^{\star}$ can be found from the obvious formal recipe

\be
\kappa_{c1}^{\star}=\left.\frac{d}{d\varepsilon}\left(\frac{1}{2\pi}\int_{0}^{2\pi}d\phi\int_{s_{c}(\phi;\varepsilon)}^{L}ds s f_{c}^{\star}(s,\phi;\varepsilon)\right)\right|_{\varepsilon=0}.\label{135}
\ee

\noindent Using (\ref{121}), (\ref{132}), and (\ref{134}), one obtains

\be
\kappa_{c1}^{\star}=-\frac{3}{4}\ln\left(\frac{L}{s_{c0}}\right).\label{136}
\ee

\noindent Hence, 

\be
\kappa_{bc1}^{\star}(\beta)\equiv\kappa_{b1}^{\star}-\kappa_{c1}^{\star}=-\frac{3}{16}-\frac{1}{24}[w_0(\beta)-1]^2-\frac{3}{4}\ln\left(\frac{s_{c0}(\beta)}{s_1(\beta)}\right)+\mathcal{O}(L^{-2})\label{137}
\ee

In \cite{disc1} the critical value of $\beta$, below which the smooth disclination line is energetically more favourable than the singular one, has been estimated to be equal to $\beta_{\mathrm{cr}}=0.12$. In the present context of the perturbative expansion that value represents the solution of the following equation

\be
\Delta\kappa_{bc0}(\beta_{\mathrm{cr}})=0,\label{138}
\ee

\noindent where $\Delta\kappa_{bc0}(\beta)$ is the zeroth order contribution to the full difference between the free energy of the smooth disclination line and that of the singular one $\Delta\kappa_{bc}^{\Sigma}(\beta)$, where the superscript refers to the three different types of the disclination line, \textit{i.e.} $\Sigma\in\{+,-,\star\}$. The general equation defining $\beta^{\Sigma}_{\mathrm{cr}}$ has the following form

\be
\Delta\kappa_{bc}^{\Sigma}(\beta_{\mathrm{cr}}^{\Sigma})=0,\label{139}
\ee
  
\noindent where $\Delta\kappa_{bc}^{\Sigma}(\beta)$ as well as  $\beta_{\mathrm{kr}}^{\Sigma}$ should be replaced by their perturbative expansions in $\varepsilon$

\beq
\Delta\kappa_{bc}^{\Sigma}(\beta)&=&\Delta\kappa_{bc0}^{\Sigma}(\beta)+\varepsilon\Delta\kappa_{bc1}^{\Sigma}(\beta)+\varepsilon^2\Delta\kappa_{bc2}^{\Sigma}(\beta)+\ldots,\label{140} \\
\beta_{\mathrm{kr}}^{\Sigma}&=&\beta^{\Sigma}_0+\varepsilon\beta^{\Sigma}_1+\varepsilon^2\beta^{\Sigma}_2+\ldots,\label{141}
\eeq

\noindent where, in accord with (\ref{138}), $\beta^{\Sigma}_{0}=\beta_{\mathrm{cr}}=0.12$. From the first order of the expansion of Eq.(\ref{139}) one obtains the formula for the first order correction to $\beta^{\Sigma}_{\mathrm{cr}}$

\be
\beta_1^{\Sigma}=-\left(\left.\frac{d\Delta\kappa_{bc0}^{\Sigma}}{d\beta}\right|_{\beta=\beta_0^\Sigma}\right)^{-1}\Delta\kappa_{bc1}^{\Sigma}(\beta_0^{\Sigma}).\label{142}
\ee

\noindent The value of the derivative appearing within the parentheses can be easily estimated with the help of the results obtained in \cite{disc1}, see Fig.4. therein. The rough estimation gives

\be
\left.\frac{d\Delta\kappa_{bc0}^{\Sigma}}{d\beta}\right|_{\beta=\beta_0^\Sigma}\approx 5.1.\label{143}
\ee

\noindent Using that estimation and the relevant results from Subsection \textbf{3.3} in \cite{disc1}, one obtains

\beq
\beta_1^{+}&\approx& 0.006, \nonumber \\
\beta_1^{-}&\approx& 0.006, \label{144} \\
\beta_1^{\star}&\approx& 0.02. \nonumber
\eeq

\noindent For each type of the disclination line the first order correction to $\beta^{\Sigma}_{\mathrm{cr}}$ is positive, which means that in each case the perturbation causes the stability area of the smooth disclination line to expand if $L_2>0$ or to shrink otherwise. Yet, even for the twist disclination line in MBBA the relevant correction is too weak to stabilize the smooth configuration; the corrected value of $\beta^{\star}_{\mathrm{cr}}$ still does not exceed the value $\beta_{m}=0.174$ which estimates the nematic-crystalline state boundary. 

\newpage

\section{Final remarks}\label{fr}

Let us recapitulate the main results of the paper.

\begin{enumerate}

\item The perturbative scheme presented here represents, to the best of our knowledge, the first analytical attempt to go 
beyond the $L_2=0$ approximation for the disclination line in the Landau-de Gennes model. In \cite{schop} only the 
numerical analysis of the problem has been carried out. The idea of the scheme rests upon the crucial assumption that the 
relevant solution can be written as a power series in the perturbative parameter $\varepsilon$.  We are not able to give 
a proof of convergence of this series. 
We have obtained the infinite system of equations, of which only the zeroth order one is nonlinear. The linear operator 
appearing in the equations for the consecutive corrections has exactly the same form once the zeroth order solution has 
been chosen. An essential part in the scheme play the zero modes whose existence is due to the symmetries of the 
unperturbed system. Mathematically, they represent eigenfunctions of the aforementioned linear operator belonging to the 
eigenvalue zero. Via consistency conditions they have been shown to select those solutions of the unperturbed system which 
are 'matching' the perturbation.

\item With the help of the Fourier analysis of the corrections the problem of solving the rather complex linear partial 
differential equations (\ref{27}) has been reduced to that of solving a system of linear ordinary differential equations 
(\ref{98}). The fact that just in the first order the number of $P$-functions (see Appendix) for, say, the twist 
disclination line amounts to 15, and will grow as one passes on to the higher orders, renders the scheme quite impractical 
in the second and higher orders. Yet, it has to be stressed that it does by no means represent any flaw of the scheme but 
is a pure consequence of the nontrivial structure of the order parameter for a disclination line in the Landau-de Gennes 
model. Moreover, one has to remember that the first order correction to the order parameter field is all one needs to 
calculate the second order correction to the free energy, which in view of the fact the the model \textit{per se} is not 
exact provides a satisfactory approximation. Another important aspect of the scheme is that it represents an approach 
complementary with purely numerical calculations \cite{schop}.

\item The scheme posseses yet another advantage. It seems that it can be applied to higher order terms in the elastic 
free energy density expansion. What is important, they would not affect the form of the linear opeartor 
$\hat{\mathcal{L}}$. Thus, the whole structure of the scheme would remain intact. Certainly, the inhomogeneous terms 
$\hat{\mathcal{N}}_k$ would be affected, and probably some alterations of the set of the unperturbed solutions cosistent 
with the higher order term in question would appear.    

\end{enumerate}

\section{Acknowledgement}
This work is supported in part by ESF Programme "Coslab". 

\section{Appendix -- the first order corrections to the order parameter}\label{ap}

\subsection{The wedge disclination line case}\label{ap1}

Let us first find the form of the correction $\hat{q}^{\sigma}_{1}$. Using (\ref{25}), (\ref{31}), (\ref{65}), (\ref{94}--\ref{97}) for $k=1$, we obtain

\be
\begin{array}{c}
\begin{array}{cc}
\vec{N}^{\sigma 0}_{1(1)}=\left(\begin{array}{c}
\frac{2}{3\sqrt{6}}\left(S^{\prime\prime}+\frac{1}{s}S^{\prime}\right) \\
0 \\
0 \\
\end{array}\right), & 
\vec{N}^{\prime\sigma 1}_{1(1)}=-\frac{1+\sigma}{2}\left(\begin{array}{c}
\frac{1}{\sqrt{6}}\left(R^{\prime\prime}+\frac{1}{s}R^{\prime}-\frac{1}{s^2}R\right) \\
0 \\
\frac{1}{3}\left(S^{\prime\prime}-\frac{1}{s}S^{\prime}\right) \\
0 \\
0 \\
0 
\end{array}\right)
\end{array}, \\
\begin{array}{cc} 
\hspace{-1 cm}\vec{N}^{\prime\sigma 3}_{1(1)}=-\frac{1-\sigma}{2}\left(\begin{array}{c}
\frac{1}{\sqrt{6}}\left(R^{\prime\prime}-\frac{3}{s}R^{\prime}+\frac{3}{s^2}R\right) \\
0 \\
\frac{1}{3}\left(S^{\prime\prime}-\frac{1}{s}S^{\prime}\right) \\
0 \\
0 \\
0
\end{array}\right), & 
\hspace{1 cm}\vec{N}^{\prime\sigma l}_{1(1)}=\vec{0}\ \ \mathrm{for}\ \  l\neq 1,3,
\end{array}
\end{array}\label{A1a}
\ee

\be
\begin{array}{cc}
\vec{N}^{\sigma 0}_{1(2)}=\vec{0}, & 
\hspace{1 cm}\vec{N}^{\prime\sigma l}_{1(2)}=\vec{0}\ \ \mathrm{for\ all}\ \  l\neq 0,
\end{array}\label{A1b}
\ee

\noindent where $R^{\prime\prime}, S^{\prime\prime}$ and $R^{\prime}, S^{\prime}$ denote the second and first derivatives with respect to $s$, respectively. For the nonvanishing components of $\vec{q}^{\ \sigma 0}_{1(1)}$, $\vec{q}^{\ \prime\sigma l}_{1(1)}$, $\vec{q}^{\ \sigma 0}_{1(2)}$ and $\vec{q}^{\ \prime\sigma l}_{1(2)}$, we accept the following notation

\be
\begin{array}{cc}
\begin{array}{l}
\left\{\begin{array}{l}
P^{+}_1=\frac{1}{\sqrt{6}}v^{+ 0}_{11}, \\
P^{+}_2=\frac{1}{\sqrt{2}}v^{+ 0}_{12}, 
\end{array}\right. \\
\left\{\begin{array}{l}
P^{+}_3=\frac{1}{\sqrt{6}}v^{\prime + 1}_{11}, \\
P^{+}_4=\frac{1}{2}v^{\prime + 1}_{12}, \\
P^{+}_5=\frac{1}{2}v^{\prime + 1}_{13},
\end{array}\right.
\end{array} \hspace{0.5 cm} & \hspace{0.5 cm}
\begin{array}{l}
\left\{\begin{array}{l}
P^{-}_1=\frac{1}{\sqrt{6}}v^{- 0}_{11}, \\
P^{-}_2=\frac{1}{\sqrt{2}}v^{- 0}_{12}, 
\end{array}\right.\\
\left\{\begin{array}{l}
P^{-}_3=\frac{1}{\sqrt{6}}v^{\prime - 3}_{11}, \\
P^{-}_4=\frac{1}{2}v^{\prime - 3}_{12}, \\
P^{-}_5=\frac{1}{2}v^{\prime - 3}_{13}.
\end{array}\right.
\end{array}
\end{array}\label{A2}
\ee

\noindent Now, using (\ref{73}--\ref{74}), (\ref{98}--\ref{99}), (\ref{A1a}--\ref{A1b}), one obtains

\be
\left\{\begin{array}{c}
P^{+ \prime\prime}_1+\frac{1}{s}P^{+ \prime}_{1}+\left[\frac{2}{3}-2\beta S-\frac{1}{2}(1+3\beta)(S^2+R^2)\right]P^{+}_{1}+\frac{1}{3}R[6\beta-(1+3\beta)S]P^{+}_{2} \\
=\frac{1}{9}\left(S^{\prime\prime}+\frac{1}{s}S^{\prime}\right), \\
\mbox{} \\
P^{+ \prime\prime}_{2}+\frac{1}{s}P^{+ \prime}_{2}-\frac{1}{s^2}P^{+}_{2}+\left[\frac{2}{3}+2\beta S-\frac{1}{6}(1+3\beta)(S^2+9R^2)\right]P^{+}_{2}+ \\
+R[6\beta-(1+3\beta)S]P^{+}_{1}=0,
\end{array}\right.\label{A4}
\ee

\be
\hspace{-3.5 cm}\left\{\begin{array}{c}
P^{+ \prime\prime}_{3}+\frac{1}{s}P^{+\prime}_{3}-\frac{1}{s^2}P^{+}_{3}+\left[\frac{2}{3}-2\beta S-\frac{1}{2}(1+3\beta)(S^2+R^2)\right]P^{+}_{3}+ \\ 
+\frac{1}{3}R[6\beta-(1+3\beta)S]\left(P^{+}_{4}+P^{+}_{5}\right)= 
 -\frac{1}{6}\left(R^{\prime\prime}+\frac{1}{s}R^{\prime}-\frac{1}{s^2}R\right), \\
\mbox{} \\
P^{+\prime\prime}_{4}+\frac{1}{s}P^{+\prime}_{4}+\left[\frac{2}{3}+2\beta S-\frac{1}{6}(1+3\beta)(S^2+6R^2)\right]P^{+}_{4}+ \\ 
+\frac{1}{2}R[6\beta-(1+3\beta)S]P^{+}_{3}-\frac{1}{2}(1+3\beta)R^2P^{+}_{5}=0, \\
\mbox{} \\
P^{+\prime\prime}_{5}+\frac{1}{s}P^{+\prime}_{5}-\frac{4}{s^2}P^{+}_{5}+\left[\frac{2}{3}+2\beta S-\frac{1}{6}(1+3\beta)(S^2+6R^2)\right]P^{+}_{5}+ \\ 
+\frac{1}{2}R[6\beta-(1+3\beta)S]P^{+}_{3}-\frac{1}{2}(1+3\beta)R^2P^{+}_{4}=
-\frac{1}{6}\left( S^{\prime\prime}-\frac{1}{s}S^{\prime}\right),
\end{array}\right.\label{A5} 
\ee

\be
\left\{\begin{array}{c}
P^{- \prime\prime}_1+\frac{1}{s}P^{- \prime}_{1}+\left[\frac{2}{3}-2\beta S-\frac{1}{2}(1+3\beta)(S^2+R^2)\right]P^{-}_{1}+\frac{1}{3}R[6\beta-(1+3\beta)S]P^{-}_{2} \\
=\frac{1}{9}\left(S^{\prime\prime}+\frac{1}{s}S^{\prime}\right), \\
\mbox{} \\
P^{- \prime\prime}_{2}+\frac{1}{s}P^{+ \prime}_{2}-\frac{1}{s^2}P^{-}_{2}+\left[\frac{2}{3}+2\beta S-\frac{1}{6}(1+3\beta)(S^2+9R^2)\right]P^{-}_{2}+ \\
+R[6\beta-(1+3\beta)S]P^{-}_{1}=0,
\end{array}\right.\label{A6}
\ee

\be
\hspace{-3.5 cm}\left\{\begin{array}{c}
P^{-\prime\prime}_{3}+\frac{1}{s}P^{-\prime}_{3}-\frac{9}{s^2}P^{-}_{3}+\left[\frac{2}{3}-2\beta S-\frac{1}{2}(1+3\beta)(S^2+R^2)\right]P^{-}_{3}+\\
+\frac{1}{3}R[6\beta-(1+3\beta)S]\left(P^{-}_{4}+P^{-}_{5}\right)=-\frac{1}{6}\left(R^{\prime\prime}-\frac{3}{s}R^{\prime}+\frac{3}{s^2}R\right), \\ 
\mbox{} \\
P^{-\prime\prime}_{4}+\frac{1}{s}P^{-\prime}_{4}-\frac{16}{s^2}P^{-}_{4}+\left[\frac{2}{3}+2\beta S-\frac{1}{6}(1+3\beta)(S^2+6R^2)\right]P^{-}_{4}+ \\
+\frac{1}{2}R[6\beta-(1+3\beta)S]P^{-}_{3}-\frac{1}{2}(1+3\beta)R^2P^{-}_{5}=0, \\ 
\mbox{} \\
P^{-\prime\prime}_{5}+\frac{1}{s}P^{-\prime}_{5}-\frac{4}{s^2}P^{-}_{5}+\left[\frac{2}{3}+2\beta S-\frac{1}{6}(1+3\beta)(S^2+R^2)\right]P^{-}_{5}+ \\
+\frac{1}{2}R[6\beta-(1+3\beta)S]P^{-}_{3}-\frac{1}{2}(1+3\beta)R^2P^{-}_{4}=-\frac{1}{6}\left(S^{\prime\prime}-\frac{1}{s}S^{\prime}\right).
\end{array}\right.\label{A7}
\ee

\noindent Note that Eqs. (\ref{A4}) and (\ref{A6}) for $P^{\pm}_1(s)$ and $P^{\pm}_2(s)$ are identical. With the help of the formulae (\ref{65}), (\ref{69}--\ref{72}), (\ref{80}), (\ref{A2}) one arrives at the final form of the corrections 

\be
\hat{q}^{+}_{1}=\left(\begin{array}{ccc}
\begin{array}{l} P^{+}_{1}+P^{+}_{4}+ \\ 
+\left(P^{+}_{3}+P^{+}_{2}\right)\cos\phi+ \\
+ P^{+}_{5}\cos 2\phi\end{array} & 
\begin{array}{l} P^{+}_{2}\sin\phi+ \\ 
+P^{+}_{5}\sin 2\phi \end{array}& 0 \\ 
\mbox{} & \mbox{} & \mbox{} \\
\begin{array}{l} P^{+}_{2}\sin\phi+ \\
+P^{+}_{5}\sin 2\phi \end{array} & 
\begin{array}{l} P^{+}_{1}-P^{+}_{4}+ \\
+\left(P^{+}_{3}-P^{+}_{2}\right)\cos\phi- \\
-P^{+}_{5}\cos 2\phi \end{array} & 0 \\
\mbox{} & \mbox{} & \mbox{} \\
0 & 0 & -2\left(P^{+}_{1}+P^{+}_{3}\cos\phi\right) 
\end{array}\right),\label{A8}
\ee

\be
\hspace{-2.3 cm}\hat{q}^{-}_{1}=\left(\begin{array}{ccc}
\begin{array}{l} P^{-}_{1}+P^{-}_{2}\cos\phi+ \\ 
+P^{-}_{5}\cos 2\phi + \\ 
+P^{-}_{3}\cos 3\phi \\
+ P^{-}_{4}\cos 4\phi\end{array} & 
\begin{array}{l} -P^{-}_{2}\sin\phi+ \\ 
+P^{-}_{5}\sin 2\phi- \\
-P^{-}_{4}\sin 4\phi\end{array}& 0 \\ 
\mbox{} & \mbox{} & \mbox{} \\
\begin{array}{l} -P^{-}_{2}\sin\phi+ \\
+P^{-}_{5}\sin 2\phi- \\
-P^{-}_{4}\sin 4\phi \end{array} & 
\begin{array}{l} P^{-}_{1}-P^{-}_{2}\cos\phi+ \\
-P^{-}_{5}\cos 2\phi+ \\
+P^{-}_{3}\cos 3\phi- \\
-P^{-}_{4}\cos 4\phi \end{array} & 0 \\
\mbox{} & \mbox{} & \mbox{} \\
0 & 0 & -2\left(P^{-}_{1}+P^{-}_{3}\cos 3\phi\right) 
\end{array}\right).\label{A9}
\ee

Our considerations would be obviously far from complete if they lacked any attention to the relevant boundary conditions. From the physical requirement that both the corrections should be continuous at $s=0$, one obtains the following boundary conditions for $P^{\pm}_{i}(s)$ ($i=1,2,\ldots,5$) 

\be
\begin{array}{lll}
\left\{\begin{array}{l}
P^{+}_1(0)=c^{+}_1, \\
P^{+}_2(0)=0, \\
P^{+}_3(0)=0, \\
P^{+}_4(0)=c^{+}_4, \\
P^{+}_5(0)=0, 
\end{array}\right. &
\hspace{2 cm} \mbox{} &
\left\{\begin{array}{l}
P^{-}_1(0)=c^{-}_1, \\
P^{-}_2(0)=0, \\
P^{-}_3(0)=0, \\
P^{-}_4(0)=0, \\
P^{-}_5(0)=0,
\end{array}\right.
\end{array}\label{A10}
\ee

\noindent where $c^{\pm}_{1}$ and $c^{\pm}_{4}$ are constants to be specified in the course of the corresponding numerical calculations. The boundary conditions for $s\rightarrow\infty$ are fully determined by another understandable requirement that the corrections should be finite. Namely, in the limit $s\rightarrow\infty$ functions $P^{\pm}_{i}(s)$ ($i=1,2,\ldots,5$) can be expanded in the following series

\be
P^{\pm}_{i}(s)=\sum_{l=0}^{\infty} P^{\pm l}_{i}\left(\frac{1}{s}\right)^{2l}, \hspace{1 cm} \mathrm{for}\ \ \  i=1,2,\ldots,5.\label{A11} 
\ee

\noindent In the above sum all odd powers of $\frac{1}{s}$ have been omitted. These terms vanish because the relevant expansions for $R(s)$ and $S(s)$ contain exclusively even powers of $\frac{1}{s}$, see \cite{disc1}. It manifests itself through Eqs.(\ref{A4}--\ref{A7}) being invariant with respect to the formal transformation $s\rightarrow -s$. For the sake of the argument to come, let us quote here the first two expansion coefficients for $R$ and $S$

\be
\begin{array}{ll}
R_{0}=1, \hspace{3 cm} & R_{2}=-\frac{1+15\beta}{12\beta(2+3\beta)}, \\
\mbox{} &  \mbox{} \\
S_{0}=1, \hspace{3 cm} & S_{2}=\frac{1-3\beta}{4\beta(2+3\beta)}.
\end{array}\label{A12}
\ee

\noindent Inserting (\ref{A11}) together with the appropriate series for $R$ and $S$ into Eqs. (\ref{A4}--\ref{A7}), one obtains the following system of equations for the expansion coefficients $P^{\pm 0}_{i}$

\be
\left\{\begin{array}{l}
(1+15\beta)P^{\pm 0}_1+(1-3\beta)P^{\pm 0}_2=0, \\
(1-3\beta)P^{\pm 0}_1+(1+3\beta)P^{\pm 0}_2=0, 
\end{array}\right.\label{A13}
\ee

\be
\left\{\begin{array}{l}
(1+15\beta)P^{\pm 0}_3+(1-3\beta)P^{\pm 0}_4+(1-3\beta)P^{\pm 0}_5=0, \\
(1-3\beta)P^{\pm 0}_3+(1+3\beta)P^{\pm 0}_4+(1+3\beta)P^{\pm 0}_5=0, \\
(1-3\beta)P^{\pm 0}_3+(1+3\beta)P^{\pm 0}_4+(1+3\beta)P^{\pm 0}_5. 
\end{array}\right.\label{A14}
\ee 

\noindent As (\ref{A13}) is the Cramer system, we get $P^{\pm 0}_1=P^{\pm 0}_2=0$. In (\ref{A14}) the two last equations are identical, so the system does not determine the values of coefficients $P^{\pm 0}_{i}$ ($i=4,5$) uniquelly, but it is equivalent to the following system 

\be
P^{\pm 0}_3=0, \hspace{2 cm} P^{\pm 0}_4+P^{\pm 0}_5=0.\label{A15} 
\ee   

\noindent For unique determination of $P^{\pm 0}_{i}$ ($i=4,5$) one has to go to the next order of the asymptotic expansion. Equating the coefficients multiplying on either side of Eqs. (\ref{A4}--\ref{A7}) the term $\left(\frac{1}{s}\right)^2$ and taking (\ref{A15}) into account, one arrives at the following systems of linear equations

\be
\left\{\begin{array}{l}
(1+15\beta)P^{+ 2}_3+(1-3\beta)P^{+ 2}_4+(1-3\beta)P^{+ 2}_5=-\frac{1}{2}, \\
\mbox{} \\
\begin{array}{l}
(1-3\beta)P^{+ 2}_3+ (1+3\beta)P^{+ 2}_4+(1+3\beta)P^{+ 2}_5= \\
2\left[2\beta S_2-\frac{1}{3}(1+3\beta)(S_2+6R_2)\right]P^{+ 0}_4-2(1+3\beta)R_2P^{+ 0}_5,
\end{array} \\
\mbox{} \\
\begin{array}{l}
(1-3\beta)P^{+ 2}_3+ (1+3\beta)P^{+ 2}_4+(1+3\beta)P^{+ 2}_5=-8P^{+ 0}_5+ \\
+2\left[2\beta S_2-\frac{1}{3}(1+3\beta)(S_2+6R_2)\right]P^{+ 0}_5-2(1+3\beta)R_2P^{+ 0}_4,
\end{array} 
\end{array}\right.\label{A16}
\ee

\be
\left\{\begin{array}{l}
(1+15\beta)P^{- 2}_3+(1-3\beta)P^{- 2}_4+(1-3\beta)P^{- 2}_5=\frac{3}{2}, \\
\mbox{} \\
\begin{array}{l}
(1-3\beta)P^{- 2}_3+ (1+3\beta)P^{- 2}_4+(1+3\beta)P^{- 2}_5=-32P^{- 0}_4+ \\
+2\left[2\beta S_2-\frac{1}{3}(1+3\beta)(S_2+6R_2)\right]P^{- 0}_4-2(1+3\beta)R_2P^{- 0}_5, 
\end{array} \\
\mbox{} \\
\begin{array}{l}
(1-3\beta)P^{- 2}_3+ (1+3\beta)P^{- 2}_4+(1+3\beta)P^{- 2}_5=-8P^{- 0}_5+ \\
+2\left[2\beta S_2-\frac{1}{3}(1+3\beta)(S_2+6R_2)\right]P^{- 0}_5-2(1+3\beta)R_2P^{- 0}_4.
\end{array} 
\end{array}\right.\label{A17}
\ee

\noindent Again, the l.h.s.'s of the two last equations in (\ref{A16}) and (\ref{A17}) are identical, whereas, in general, their r.h.s.'s differ. Hence, for the systems to be consistent one has to demand that their relevant r.h.s.'s should be equal. This requirement leads to the following system

\be
\left\{\begin{array}{l}
P^{+ 0}_4+3P^{+ 0}_5=0, \\
5P^{- 0}_4-P^{- 0}_5=0,
\end{array}\right. ,\label{A19}
\ee 

\noindent which together with the second equation in (\ref{A15}) give uniquely$P^{\pm 0}_4=P^{\pm 0}_5=0$. To conclude, the consistent boundary conditions for $P^{\pm}_{i}(s)$ ($i=1,2,\ldots,5$) have the following form

\be
P^{\pm}_i({\infty})=0, \hspace{1.5 cm}\ \ \ i=1,2\ldots,5.\label{A20}
\ee

\subsection{The twist disclination line case}\label{ap2}

For the non-vanishing components of $\hat{\tilde{\mathcal{N}}}^{\perp}_{1}$, using (\ref{31}), (\ref{65}), (\ref{94}--\ref{97}), (\ref{104}), and (\ref{105}) with $k=1$, one obtains

\be
\vec{\tilde{N}}^{\perp 0}_{1(1)}=\left(\begin{array}{c}
\tilde{V}^{\perp 0}_{11} \\
\tilde{V}^{\perp 0}_{12} \\
\tilde{V}^{\perp 0}_{13}
\end{array}\right)=
\left(\begin{array}{c}
\frac{1}{6\sqrt{6}}\left(S^{\prime\prime}+\frac{1}{s}S^{\prime}\right) \\
0 \\
0 
\end{array}\right),\label{A21}
\ee

\be
\vec{\tilde{N}}^{\prime\perp 1}_{1(1)}=\left(\begin{array}{c}
\tilde{V}^{\prime\perp 1}_{11} \\
\tilde{V}^{\prime\perp 1}_{12} \\
\tilde{V}^{\prime\perp 1}_{13} \\   
\tilde{U}^{\prime\perp 1}_{11} \\
\tilde{U}^{\prime\perp 1}_{12} \\
\tilde{U}^{\prime\perp 1}_{13}
\end{array}\right)=
\left(\begin{array}{c}
\frac{1}{4\sqrt{6}}\cos 2\Psi\left(R^{\prime\prime}+\frac{1}{s}R^{\prime}-\frac{1}{s^2}R\right) \\
\frac{1}{6}\cos 2\Psi\left(S^{\prime\prime}+\frac{1}{s}S^{\prime}\right) \\
-\frac{1}{12}\cos 2\Psi\left(S^{\prime\prime}-\frac{1}{s}S^{\prime}\right) \\
-\frac{3}{4\sqrt{6}}\sin 2\Psi\left(R^{\prime\prime}+\frac{1}{s}R^{\prime}-\frac{1}{s^2}R\right) \\
-\frac{1}{6}\sin 2\Psi\left(S^{\prime\prime}+\frac{1}{s}S^{\prime}\right) \\
-\frac{1}{12}\sin 2\Psi\left(S^{\prime\prime}-\frac{1}{s}S^{\prime}\right)
\end{array}\right),\label{A22}
\ee

\be
\vec{\tilde{N}}^{\prime\perp 2}_{1(1)}=\left(\begin{array}{c}
\tilde{V}^{\prime\perp 2}_{11} \\
\tilde{V}^{\prime\perp 2}_{12} \\
\tilde{V}^{\prime\perp 2}_{13} \\   
\tilde{U}^{\prime\perp 2}_{11} \\
\tilde{U}^{\prime\perp 2}_{12} \\
\tilde{U}^{\prime\perp 2}_{13}
\end{array}\right)=
\left(\begin{array}{c}
-\frac{1}{2\sqrt{6}}\left(S^{\prime\prime}-\frac{1}{s}S^{\prime}\right) \\
\frac{1}{4}\left(R^{\prime\prime}+\frac{1}{s}R^{\prime}-\frac{1}{s^2}R\right) \\
\frac{1}{4}\left(R^{\prime\prime}-\frac{3}{s}R^{\prime}+\frac{3}{s^2}R\right) \\
0 \\
0 \\
0 
\end{array}\right),\label{A23}
\ee

\be
\vec{\tilde{N}}^{\prime\perp 3}_{1(1)}=\left(\begin{array}{c}
\tilde{V}^{\prime\perp 3}_{11} \\
\tilde{V}^{\prime\perp 3}_{12} \\
\tilde{V}^{\prime\perp 3}_{13} \\   
\tilde{U}^{\prime\perp 3}_{11} \\
\tilde{U}^{\prime\perp 3}_{12} \\
\tilde{U}^{\prime\perp 3}_{13}
\end{array}\right)=
\left(\begin{array}{c}
-\frac{1}{4\sqrt{6}}\cos 2\Psi\left(R^{\prime\prime}-\frac{3}{s}R^{\prime}+\frac{3}{s^2}R\right) \\
-\frac{1}{12}\cos 2\Psi\left(S^{\prime\prime}-\frac{1}{s}S^{\prime}\right) \\
0 \\
\frac{1}{4\sqrt{6}}\sin 2\Psi\left(R^{\prime\prime}-\frac{3}{s}R^{\prime}+\frac{3}{s^2}R\right) \\
\frac{1}{12}\sin 2\Psi\left(S^{\prime\prime}-\frac{1}{s}S^{\prime}\right) \\
0
\end{array}\right),\label{A24}
\ee

\be
\vec{\tilde{N}}^{\prime\perp 2}_{1(2)}=\left(\begin{array}{c}
\tilde{V}^{\prime\perp 2}_{14} \\
\tilde{V}^{\prime\perp 2}_{15} \\
\tilde{U}^{\prime\perp 2}_{14} \\
\tilde{U}^{\prime\perp 2}_{15} \\
\end{array}\right)=
\left(\begin{array}{c}
-\frac{1}{4}\cos\Psi\left(R^{\prime\prime}+\frac{1}{s}R^{\prime}-\frac{1}{s^2}R\right) \\
\frac{1}{12}\cos\Psi\left(S^{\prime\prime}-\frac{1}{s}S^{\prime}\right) \\
-\frac{1}{4}\sin\Psi\left(R^{\prime\prime}+\frac{1}{s}R^{\prime}-\frac{1}{s^2}R\right) \\
\frac{1}{12}\sin\Psi\left(S^{\prime\prime}-\frac{1}{s}S^{\prime}\right)
\end{array}\right),\label{A25}
\ee

\be
\vec{\tilde{N}}^{\prime\perp 2}_{1(2)}=\left(\begin{array}{c}
\tilde{V}^{\prime\perp 3}_{14} \\
\tilde{V}^{\prime\perp 3}_{15} \\
\tilde{U}^{\prime\perp 3}_{14} \\
\tilde{U}^{\prime\perp 3}_{15} \\
\end{array}\right)=
\left(\begin{array}{c}
-\frac{1}{12}\cos\Psi\left(S^{\prime\prime}-\frac{1}{s}S^{\prime}\right) \\
\frac{1}{4}\cos\Psi\left(R^{\prime\prime}-\frac{3}{s}R^{\prime}+\frac{3}{s^2}R\right) \\
\frac{1}{12}\sin\Psi\left(S^{\prime\prime}-\frac{1}{s}S^{\prime}\right) \\
-\frac{1}{4}\sin\Psi\left(R^{\prime\prime}-\frac{3}{s}R^{\prime}+\frac{3}{s^2}R\right)
\end{array}\right),\label{A26}
\ee
 
\noindent where $\Psi$ takes on one of the four values $\{0,\frac{\pi}{4},\frac{\pi}{2},\frac{3\pi}{4}\}$ (for each configuration considered here $\Theta=\frac{\pi}{2}$). One should remember that in the above formulae the symbol $\mbox{}^{\prime}$ denotes the differentiation with respect to $s$ exclusivly for the functions $R$ or $S$. Otherwise, it refers to the orthogonal transformations $\mathcal{O}^{(1)}$ and $\mathcal{O}^{(2)}$ in the submodules $\mathcal{F}^{+(1)}_{l}$ and $\mathcal{F}^{-(2)}_{l}$ ($l\neq 0$), respectively. The system of equations pertinent to the discussed case has the form

\be
\left\{\begin{array}{l}
\hat{\mathcal{L}}^{+ (1)}_{0}\vec{\tilde{q}}^{\perp 0}_{k(1)}=\vec{\tilde{N}}^{\perp 0}_{k(1)}, \\
\hat{\mathcal{L}}^{\prime + (1)}_{l}\vec{\tilde{q}}^{\perp l}_{k(1)}=\vec{\tilde{N}}^{\prime\perp l}_{k(1)} \ \ \ \ \mathrm{for}\ \ \ l\neq 0,\\ 
\hat{\mathcal{L}}^{+ (2)}_{0}\vec{\tilde{q}}^{\perp 0}_{k(2)}=\vec{\tilde{N}}^{\perp 0}_{k(2)}, \\
\hat{\mathcal{L}}^{\prime + (2)}_{l}\vec{\tilde{q}}^{\perp l}_{k(2)}=\vec{\tilde{N}}^{\prime\perp l}_{k(2)}\ \ \ \  \mathrm{for}\ \ \ l\neq 0,
\end{array}\right.\label{A27}
\ee

\noindent where

\be
\begin{array}{l}
\vec{\tilde{q}}^{\perp 0}_{k(1)}\equiv [\tilde{v}^{\perp 0}_{ka}]_{\{a=1,2,3\}}, \\
\vec{\tilde{q}}^{\ \prime\perp l}_{k(1)}\equiv [\tilde{v}^{\prime\perp l}_{ka},\tilde{u}^{\prime\perp l}_{ka}]_{\{a=1,2,3\}} \ \ \ \ \mathrm{for}\ \ \  l\neq 0, \\
\vec{\tilde{q}}^{\perp 0}_{k(2)}\equiv [\tilde{v}^{\perp 0}_{ka}]_{\{a=4,5\}}, \\
\vec{\tilde{q}}^{\ \prime\perp l}_{ka}\equiv [\tilde{v}^{\prime\perp l}_{ka},\tilde{u}^{\prime\perp l}_{ka}]_{\{a=4,5\}}\ \ \ \ \ \ \mathrm{for}\ \ \ l\neq 0, \end{array}\label{A28}
\ee 

\noindent for $k=1,2,\ldots$. Specifically, for $k=1$, using (\ref{A21}--\ref{A28}), one can obtain the final forms of corrections $\hat{q}^{\perp}_{1}$ and write the analogons of Eqs. (\ref{A4}--\ref{A7}) for all the four cases. Due to the length of the resulting formulae we confine ourselves to quoting the result for one case only, \textit{i.e.} for $\Psi=0$. Let us replace the superscript $\mbox{}^\perp$ by $\mbox{}^\star$ in all quantities referring to that case. Now, for the non-vanishing components of $\vec{\tilde{q}}^{\star 0}_{1(1)}$, $\vec{\tilde{q}}^{\ \prime\star l}_{1(1)}$, $\vec{\tilde{q}}^{\star 0}_{1(2)}$ i $\vec{\tilde{q}}^{\ \prime\star l}_{1(2)}$ ($l\neq 0$) we introduce the following notation

\be
\begin{array}{ll}
\begin{array}{l}
\left\{\begin{array}{l}
P^{\star}_1=\frac{1}{\sqrt{6}}\tilde{v}^{\star 0}_{11}, \\
P^{\star}_2=\frac{1}{\sqrt{2}}\tilde{v}^{\star 0}_{12},
\end{array}\right. \\
\left\{\begin{array}{l}
P^{\star}_3=\frac{1}{\sqrt{6}}\tilde{v}^{\prime\star 1}_{11}, \\
P^{\star}_4=\frac{1}{2}\tilde{v}^{\prime\star 1}_{12}, \\
P^{\star}_5=\frac{1}{2}\tilde{v}^{\prime\star 1}_{13}, 
\end{array}\right. \\
\left\{\begin{array}{l}
P^{\star}_6=\frac{1}{\sqrt{6}}\tilde{v}^{\prime\star 2}_{11}, \\
P^{\star}_7=\frac{1}{2}\tilde{v}^{\prime\star 2}_{12}, \\
P^{\star}_8=\frac{1}{2}\tilde{v}^{\prime\star 2}_{13}, 
\end{array}\right.
\end{array}  & 
\hspace{1 cm}\begin{array}{l}
\left\{\begin{array}{l}
P^{\star}_9=\frac{1}{\sqrt{6}}\tilde{v}^{\prime\star 3}_{11}, \\
P^{\star}_{10}=\frac{1}{2}\tilde{v}^{\prime\star 3}_{12}, \\
P^{\star}_{11}=\frac{1}{2}\tilde{v}^{\prime\star 3}_{13}, 
\end{array}\right. \\
\left\{\begin{array}{l}
P^{\star}_{12}=\frac{1}{2}\tilde{v}^{\prime\star 2}_{14}, \\
P^{\star}_{13}=\frac{1}{2}\tilde{v}^{\prime\star 2}_{15}, 
\end{array}\right. \\
\left\{\begin{array}{l}
P^{\star}_{14}=\frac{1}{2}\tilde{v}^{\prime\star 3}_{14}, \\
P^{\star}_{15}=\frac{1}{2}\tilde{v}^{\prime\star 3}_{15}.
\end{array}\right.
\end{array}
\end{array}\label{A29}
\ee

\noindent Using (\ref{A21}--\ref{A29}), one arrives at the following systems of ordinary differential equations for $P^{\star}_{i}(s)$ ($i=1,2,\ldots,15$)

\be
\left\{\begin{array}{l}
\begin{array}{c}
P^{\star\prime\prime}_1+\frac{1}{s}P^{\star\prime}_1+\left[\frac{2}{3}-2\beta S-\frac{1}{2}(1+3\beta)(S^2+R^2)\right]P^{\star}_1+\frac{1}{3}R[6\beta-(1+3\beta)S]P^{\star}_2 \\
=\frac{1}{36}\left(S^{\prime\prime}+\frac{1}{s}S^{\prime}\right), 
\end{array} \\
\mbox{} \\
\begin{array}{c}
P^{\star\prime\prime}_2+\frac{1}{s}P^{\star\prime}_2-\frac{1}{s^2}P^{\star}_2+\left[\frac{2}{3}+2\beta S-\frac{1}{6}(1+3\beta)(S^2+9R^2)\right]P^{\star}_2+ \\
+R[6\beta-(1+3\beta)S]P^{\star}_1=0, 
\end{array}
\end{array}\right.\label{A30}
\ee

\be
\hspace{-3.3 cm}\left\{\begin{array}{l}
\begin{array}{c}
P^{\star\prime\prime}_{3}+\frac{1}{s}P^{\star\prime}_3-\frac{1}{s^2}P^{\star}_3+\left[\frac{2}{3}-2\beta S-\frac{1}{2}(1+3\beta)(S^2+R^2)\right]P^{\star}_3+ \\
+\frac{1}{3}R[6\beta-(1+3\beta)S]\left(P^{\star}_4+P^{\star}_5\right)=\frac{1}{24}\left(R^{\prime\prime}+\frac{1}{s}R^{\prime}-\frac{1}{s^2}R\right),
\end{array} \\
\mbox{} \\
\begin{array}{c} 
P^{\star\prime\prime}_4+\frac{1}{s}P^{\star\prime}_4+\left[\frac{2}{3}+2\beta S-\frac{1}{6}(1+3\beta)(S^2+6R^2)\right]P^{\star}_4+ \\
+\frac{1}{2}R[6\beta-(1+3\beta)S]P^{\star}_3-\frac{1}{2}(1+3\beta)R^2P^{\star}_5=\frac{1}{12}\left(S^{\prime\prime}+\frac{1}{s}S^{\prime}\right), 
\end{array} \\
\mbox{} \\
\begin{array}{c}
P^{\star\prime\prime}_{5}+\frac{1}{s}P^{\star\prime}_5-\frac{4}{s^2}P^{\star}_5+\left[\frac{2}{3}+2\beta S-\frac{1}{6}(1+3\beta)(S^2+6R^2)\right]P^{\star}_5+ \\
+\frac{1}{2}R[6\beta-(1+3\beta)S]P^{\star}_3-\frac{1}{2}(1+3\beta)R^2P^{\star}_4=-\frac{1}{24}\left(S^{\prime\prime}-\frac{1}{s}S^{\prime}\right), 
\end{array}
\end{array}\right.\label{A31}
\ee

\be
\hspace{-2.5 cm}\left\{\begin{array}{l}
\begin{array}{c}
P^{\star\prime\prime}_6+\frac{1}{s}P^{\star\prime}_6-\frac{4}{s^2}P^{\star}_6+\left[\frac{2}{3}-2\beta S-\frac{1}{2}(1+3\beta)(S^2+R^2)\right]P^{\star}_6+ \\
+\frac{1}{3}R[6\beta-(1+3\beta)S]\left(P^{\star}_7+P^{\star}_8\right)=-\frac{1}{12}\left(S^{\prime\prime}-\frac{1}{s}S^{\prime}\right), 
\end{array} \\
\mbox{} \\
\begin{array}{c}
P^{\star\prime\prime}_7+\frac{1}{s}P^{\star\prime}_7-\frac{1}{s^2}P^{\star}_7+\left[\frac{2}{3}+2\beta S-\frac{1}{6}(1+3\beta)(S^2+6R^2)\right]P^{\star}_7+ \\
+\frac{1}{2}R[6\beta-(1+3\beta)S]P^{\star}_6-\frac{1}{2}(1+3\beta)R^2P^{\star}_8=\frac{1}{8}\left(R^{\prime\prime}+\frac{1}{s}R^{\prime}-\frac{1}{s^2}R\right), 
\end{array} \\
\mbox{} \\
\begin{array}{c}
P^{\star\prime\prime}_8+\frac{1}{s}P^{\star\prime}_8-\frac{9}{s^2}P^{\star}_8+\left[\frac{2}{3}+2\beta S-\frac{1}{6}(1+3\beta)(S^2+6R^2)\right]P^{\star}_8+ \\
+\frac{1}{2}R[6\beta-(1+3\beta)S]P^{\star}_6-\frac{1}{2}(1+3\beta)R^2P^{\star}_7=\frac{1}{8}\left(R^{\prime\prime}-\frac{3}{s}R^{\prime}+\frac{3}{s^2}R\right), 
\end{array}
\end{array}\right.\label{A32}
\ee

\be
\hspace{-3.2 cm}\left\{\begin{array}{l}
\begin{array}{c}
P^{\star\prime\prime}_9+\frac{1}{s}P^{\star\prime}_9-\frac{9}{s^2}P^{\star}_9+\left[\frac{2}{3}-2\beta S-\frac{1}{2}(1+3\beta)(S^2+R^2)\right]P^{\star}_9+ \\
+\frac{1}{3}R[6\beta-(1+3\beta)S]\left(P^{\star}_{10}+P^{\star}_{11}\right)=-\frac{1}{24}\left(R^{\prime\prime}-\frac{3}{s}R^{\prime}+\frac{3}{s^2}R\right), 
\end{array} \\
\mbox{} \\
\begin{array}{c}
P^{\star\prime\prime}_{10}+\frac{1}{s}P^{\star\prime}_{10}-\frac{4}{s^2}P^{\star}_{10}+\left[\frac{2}{3}+2\beta S-\frac{1}{6}(1+3\beta)(S^2+6R^2)\right]P^{\star}_{10}+ \\
+\frac{1}{2}R[6\beta-(1+3\beta)S]P^{\star}_9-\frac{1}{2}(1+3\beta)R^2P^{\star}_{11}=-\frac{1}{24}\left(S^{\prime\prime}-\frac{1}{s}S^{\prime}\right), 
\end{array} \\
\mbox{} \\
\begin{array}{c}
P^{\star\prime\prime}_{11}+\frac{1}{s}P^{\star\prime}_{11}-\frac{16}{s^2}P^{\star}_{11}+\left[\frac{2}{3}+2\beta S-\frac{1}{6}(1+3\beta)(S^2+6R^2)\right]P^{\star}_{11}+ \\
+\frac{1}{2}R[6\beta-(1+3\beta)S]P^{\star}_9-\frac{1}{2}(1+3\beta)R^2P^{\star}_{10}=0, 
\end{array}
\end{array}\right.\label{A33}
\ee

\be
\left\{\begin{array}{l}
\begin{array}{c}
P^{\star\prime\prime}_{12}+\frac{1}{s}P^{\star\prime}_{12}-\frac{1}{s^2}P^{\star}_{12}+\left[\frac{2}{3}-\beta S-\frac{1}{6}(1+3\beta)(S^2+3R^2)\right]P^{\star}_{12}+ \\
+3\beta R P^{\star}_{13}=-\frac{1}{8}\left(R^{\prime\prime}+\frac{1}{s}R^{\prime}-\frac{1}{s^2}R\right), 
\end{array} \\
\mbox{} \\
\begin{array}{c}
P^{\star\prime\prime}_{13}+\frac{1}{s}P^{\star\prime}_{13}-\frac{4}{s^2}P^{\star}_{13}+\left[\frac{2}{3}-\beta S-\frac{1}{6}(1+3\beta)(S^2+3R^2)\right]P^{\star}_{13}+ \\
+3\beta R P^{\star}_{12}=\frac{1}{24}\left(S^{\prime\prime}-\frac{1}{s}S^{\prime}\right), 
\end{array}
\end{array}\right.\label{A34}
\ee

\be
\left\{\begin{array}{l}
\begin{array}{c}
P^{\star\prime\prime}_{14}+\frac{1}{s}P^{\star\prime}_{14}-\frac{4}{s^2}P^{\star}_{14}+\left[\frac{2}{3}-\beta S-\frac{1}{6}(1+3\beta)(S^2+3R^2)\right]P^{\star}_{14}+ \\
+3\beta R P^{\star}_{15}=-\frac{1}{24}\left(S^{\prime\prime}-\frac{1}{s}S^{\prime}\right), 
\end{array} \\
\mbox{} \\
\begin{array}{c}
P^{\star\prime\prime}_{15}+\frac{1}{s}P^{\star\prime}_{15}-\frac{9}{s^2}P^{\star}_{15}+\left[\frac{2}{3}-\beta S-\frac{1}{6}(1+3\beta)(S^2+3R^2)\right]P^{\star}_{15}+ \\
+3\beta R P^{\star}_{14}=\frac{1}{8}\left(R^{\prime\prime}-\frac{3}{s}R^{\prime}+\frac{3}{s^2}R\right). 
\end{array}
\end{array}\right.\label{A35}
\ee

\noindent Using (\ref{57}), (\ref{65}), (\ref{69}), (\ref{80}--\ref{82}), (\ref{92}--\ref{93}), (\ref{104}), and (\ref{A29}), one obtains the final form of $\hat{q}^{\star}_{1}$

\bd
\left(\!\!\begin{array}{ccc}
\begin{array}{c}
-2\left[P^{\star}_1+P^{\star}_3\cos\phi+\right. \\
\left.+P^{\star}_6\cos 2\phi+P^{\star}_9\cos 3\phi\right]
\end{array} & 
\begin{array}{c}
-P^{\star}_{12}\sin\phi+ \\
+\left(P^{\star}_{13}-P^{\star}_{14}\right)\sin 2\phi+ \\
+P^{\star}_{15}\sin 3\phi 
\end{array}& 
\begin{array}{c}
-P^{\star}_{12}\cos\phi- \\
-\left(P^{\star}_{13}+P^{\star}_{14}\right)\cos 2\phi- \\
-P^{\star}_{15}\cos 3\phi
\end{array} \\
\mbox{} \\
\begin{array}{c}
-P^{\star}_{12}\sin\phi+ \\
+\left(P^{\star}_{13}-P^{\star}_{14}\right)\sin 2\phi+ \\
+P^{\star}_{15}\sin 3\phi 
\end{array} & 
\begin{array}{c}
P^{\star}_1-P^{\star}_4+ \\
+\left(P^{\star}_3-P^{\star}_2-P^{\star}_7\right)\cos\phi+ \\
+\left(P^{\star}_6-P^{\star}_5-P^{\star}_{10}\right)\cos 2\phi+ \\
+\left(P^{\star}_9-P^{\star}_8\right)\cos 3\phi- \\
-P^{\star}_{11}\cos 4\phi
\end{array} & 
\begin{array}{c}
\left(P^{\star}_7-P^{\star}_2\right)\sin\phi+ \\
+\left(P^{\star}_{10}-P^{\star}_5\right)\sin 2\phi- \\
-P^{\star}_8\sin 3\phi-P^{\star}_{11}\sin 4\phi
\end{array} \\
\mbox{} \\
\begin{array}{c}
-P^{\star}_{12}\cos\phi- \\
-\left(P^{\star}_{13}+P^{\star}_{14}\right)\cos 2\phi- \\
-P^{\star}_{15}\cos 3\phi
\end{array} & 
\begin{array}{c}
\left(P^{\star}_7-P^{\star}_2\right)\sin\phi+ \\
+\left(P^{\star}_{10}-P^{\star}_5\right)\sin 2\phi- \\
-P^{\star}_8\sin 3\phi-P^{\star}_{11}\sin 4\phi
\end{array} & 
\begin{array}{c}
P^{\star}_1+P^{\star}_4+ \\
+\left(P^{\star}_3+P^{\star}_2+P^{\star}_7\right)\cos\phi+ \\
+\left(P^{\star}_6+P^{\star}_5+P^{\star}_{10}\right)\cos 2\phi+ \\
+\left(P^{\star}_9+P^{\star}_8\right)\cos 3\phi+ \\
+P^{\star}_{11}\cos 4\phi
\end{array}
\end{array}\!\!\right)
\ed 

\noindent In a way analogous to that described in detail in Appendix \ref{ap1}, one can obtain the consistent boundary conditions for the functions $P^{\star}_{i}(s)$ ($i=1,2,\ldots,15$). They read

\be
\begin{array}{lll}
\left\{\begin{array}{l}
P^{\star}_1(0)=c^{\star}_1, \\
P^{\star}_2(0=0, 
\end{array}\right. &
\hspace{2 cm}\mbox{} &
\left\{\begin{array}{l}
P^{\star}_1(\infty)=0, \\
P^{\star}_2(\infty)=0, 
\end{array}\right.
\end{array}\label{A37}
\ee

\be
\begin{array}{lll}
\left\{\begin{array}{l}
P^{\star}_3(0)=0, \\
P^{\star}_4(0)=c^{\star}_4,\\
P^{\star}_5(0)=0, 
\end{array}\right. &
\hspace{2 cm}\mbox{} &
\left\{\begin{array}{l}
P^{\star}_3(\infty)=0, \\
P^{\star}_4(\infty)=0, \\
P^{\star}_5(\infty)=0, 
\end{array}\right.
\end{array}\label{A38}
\ee

\be
\hspace{0.6 cm}\begin{array}{lll}
\left\{\begin{array}{l}
P^{\star}_6(0)=0, \\
P^{\star}_7(0)=0,\\
P^{\star}_8(0)=0, 
\end{array}\right. &
\hspace{2.1 cm}\mbox{} &
\left\{\begin{array}{l}
P^{\star}_6(\infty)=0, \\
P^{\star}_7(\infty)=\frac{1}{16}, \\
P^{\star}_8(\infty)=-\frac{1}{16}, 
\end{array}\right.
\end{array}\label{A39}
\ee

\be
\hspace{0.2 cm}\begin{array}{lll}
\left\{\begin{array}{l}
P^{\star}_9(0)=0, \\
P^{\star}_{10}(0)=0,\\
P^{\star}_{11}(0)=0, 
\end{array}\right. &
\hspace{2 cm}\mbox{} &
\left\{\begin{array}{l}
P^{\star}_9(\infty)=0, \\
P^{\star}_{10}(\infty)=0, \\
P^{\star}_{11}(\infty)=0, 
\end{array}\right.
\end{array}\label{A40}
\ee

\be
\hspace{0.7 cm}\begin{array}{lll}
\left\{\begin{array}{l}
P^{\star}_{12}(0)=0, \\
P^{\star}_{13}(0)=0, 
\end{array}\right. &
\hspace{2 cm}\mbox{} &
\left\{\begin{array}{l}
P^{\star}_{12}(\infty)=-\frac{1}{32}, \\
P^{\star}_{13}(\infty)=-\frac{1}{32},  
\end{array}\right.
\end{array}\label{A41}
\ee

\be
\begin{array}{lll}
\left\{\begin{array}{l}
P^{\star}_{14}(0)=0, \\
P^{\star}_{15}(0)=0, 
\end{array}\right. &
\hspace{2 cm}\mbox{} &
\left\{\begin{array}{l}
P^{\star}_{14}(\infty)=-\frac{1}{32}, \\
P^{\star}_{15}(\infty)=-\frac{1}{32},  
\end{array}\right.\label{A42}
\end{array}
\ee

\noindent where $c^{\star}_1$ and $c^{\star}_4$ are constants to be determined in the course of numerically solving Eqs. (\ref{A30},\ref{A31}).

\end{document}